\documentclass[a4paper,11pt]{article}
\usepackage{jheppub} 
\usepackage{lineno}

\title{Particle Swarm Optimization Based Analysis to Unlocking the Neutrino Mass Puzzle using $SU(2)_L \times U(1)_Y \times A_{4}\times S_2\times Z_{10} \times Z_{3}$ Flavor Symmetry}
\author[a]{M.W.~Aslam,}
\author[a]{A.A.~Zafar,}
\author[b]{M.N.~Aslam,}
\author[c]{A.A~Bhatti,}
\author[c]{T.~Hussain,}
\author[d]{M.~Iqbal,}
\author[e,*]{and Mohammed~S.~Abdo}
\affiliation[a]{Department of Physics, University of the Punjab,\\
Lahore, Pakistan}
\affiliation[b]{Department of Mathematics, Lahore Garrison University,\\
Lahore, Pakistan}
\affiliation[c]{Centre for High Energy Physics, University of the Punjab,\\
Lahore, Pakistan}
\affiliation[d]{College of Statistical Sciences, University of the Punjab,\\
Lahore, Pakistan}
\affiliation[e]{Department of Mathematics, Hodeidah University, Al-Hudaydah, P.O. Box 3114, Yemen
}
\emailAdd{msabdo@hoduniv.net.ye}

\abstract{New research has highlighted a shortfall in the Standard Model (SM) because it predicts neutrinos to have zero mass. However, recent experiments on neutrino oscillation have revealed that the majority of neutrino parameters indeed indicate their significant mass. In response, scientists are increasingly incorporating discrete symmetries alongside continuous ones for the  observed patterns of neutrino mixing. In this study, we have examined a model within $SU(2)_L \times U(1)_Y \times A_{4}\times S_2\times Z_{10} \times Z_{3}$ symmetry to estimate the neutrino masses using particle swarm optimization technique for both mass hierarchy of neutrino. This model employed a hybrid seesaw mechanism, a combination of seesaw mechanism of type-I and type-II, to establish the effective Majorana neutrino mass matrix. After calculating the mass eigenvalues and lepton mixing matrix upto second order perturbation theory in this framework, this study seeks to investigate the scalar potential for vacuum expectation values (VEVs), optimize the parameters, $U_{PMNS}$ matrix, neutrino masses: ${m_{1}^{\prime}}^{(N)}(upper)=4.0000 \times 10^{-2}\ eV$, ${m_{2}^{\prime}}^{(N)}(upper)=4.0000 \times 10^{-2}\ eV$,  
${m_{3}^{\prime}}^{(N)}(upper)=4.0000 \times 10^{-2}\ eV$, ${m_{1}^{\prime}}^{(I)}(upper)=3.8628\times 10^{-2}\ eV$,
${m_{2}^{\prime}}^{(I)}(upper)=4.0548\times 10^{-2}\ eV$,
${m_{3}^{\prime}}^{(I)}(upper)=3.8532\times 10^{-2}\ eV$,
${m_{1}^{\prime}}^{(N)}(lower)=2.0000 \times 10^{-2}\ eV$, 
${m_{2}^{\prime}}^{(N)}(lower)=2.0000 \times 10^{-2}\ eV$, 
${m_{3}^{\prime}}^{(N)}(lower)=2.0000 \times 10^{-2}\ eV$, 
${m_{1}^{\prime}}^{(I)}(lower)=1.1049\times 10^{-2}\ eV$,
${m_{2}^{\prime}}^{(I)}(lower)=3.9298\times 10^{-2}\ eV$ and
${m_{3}^{\prime}}^{(I)}(lower)=9.6381\times 10^{-3}\ eV$, effective neutrino mass parameters: 
$\langle {m_{ee}} \rangle^{N}(upper)=40.0050 \ meV$, $\langle {m_{\beta}} \rangle^{N}(upper)=40.0025\ meV$, $\langle {m_{ee}} \rangle^{I}(upper)=39.2181\ meV$, $\langle {m_{\beta}} \rangle^{I}(upper)=39.2257\ meV$, $\langle {m_{ee}} \rangle^{N}(lower)=20.0024\ meV$, $\langle {m_{\beta}} \rangle^{N}(lower)=20.0012\ meV$, $\langle {m_{ee}} \rangle^{I}(lower)=19.6608\ meV$, $\langle {m_{\beta}} \rangle^{I}(lower)=23.5908\ meV$, are predicted for both mass hierarchy through particle swarm optimization (PSO), showing strong agreement with recent experimental findings. The Dirac CP-violating phase $\delta$ is measured to be $-\pi/2$.}
\begin{document}
\maketitle
\flushbottom

\section{Introduction}
\label{sec:intro}
Known as the "ghost particles" of the universe, neutrinos have long intrigued the interest of cosmologists and physicists alike. Despite being among the most prevalent particles in the universe, these elementary particles (which are electrically neutral and almost massless) interact with matter very weakly, which makes them notoriously difficult to detect. In 1930, Wolfgang Pauli postulated the existence of neutrinos as a possible explanation for the violation of energy conservation observed in beta decay, they were only ever considered theoretical particles. Frederick Reines and Clyde Cowan eventually detected neutrinos in 1956 \cite{cowan1956detection}. Neutrinos, in spite of their spectral appearance, are essential to the understanding of fundamental physics and the universe's evolution. Many astrophysical processes, such as nuclear fusion in stars \cite{raffelt2008neutrinos}, supernova explosions \cite{burrows1990neutrinos,tamborra2018neutrinos,burrows2000neutrinos,cooperstein1988neutrinos,herant1997neutrinos}, and even the Big Bang itself \cite{sarkar2003neutrinos,steigman2005neutrinos}, produce them. Neutrinos are also essential for solving some of the most significant mysteries in cosmology and particle physics, including the properties of the elusive Higgs boson \cite{chakraborty2014higgs}, the nature of dark matter \cite{blennow2013probing,agarwalla2011neutrino,dodelson1994sterile,bertone2018new,hooper2008strategies}, and the universe's imbalance between matter and antimatter \cite{t2k2020constraint}. One of the central puzzles surrounding neutrinos is their masses \cite{weinheimer2013neutrino}. Neutrinos were originally thought to be massless in accordance with the SM of particle physics \cite{bilenky2018introduction}. However, experiments in the late 20th and early 21st centuries, such as those conducted by the Super-Kamiokande \cite{kamiokande1998evidence},  KamLAND \cite{eguchi2003first}, K2K \cite{ahn2006measurement}, Fermilab-MINOS \cite{michael2006observation}, Sudbury Neutrino Observatory collaborations \cite{kamiokande1998evidence} and  CERN-OPERA \cite{sh2009opera} provided irrefutable evidence that neutrinos oscillate between different flavors (electron, muon, and tau), a phenomenon that can only occur if they possess non-zero masses. This discovery fundamentally challenged our understanding of neutrinos and underscored the need for new theoretical frameworks beyond the SM.

Researchers are looking in detail into seesaw frameworks, particularly type-I \cite{cai2018lepton, mohapatra2005seesaw, king2013neutrino, mohapatra2006neutrino, king2003neutrino} and type-II \cite{melfo2012type, perez2008neutrino, cheng1980neutrino}, aside from  several other methods to explain small neutrino masses. Majorana and Dirac mass terms derive from the introduction of extra right-handed neutrinos in SM in type I. Majorana mass terms derive from the introduction of heavy $SU(2)_L$ triplet in SM in type-II. A hybrid seesaw mechanism \cite{cai2018lepton, akhmedov2007interplay, wong2022tree} has been proposed for improved mass suppression and new mixing patterns by combining type-I and type-II. With this hybrid technique, one may explore various lepton mixing scenarios and generate effective Majorana neutrino mass matrices.

Considering the recent discovery of non-zero, small neutrino masses in multiple neutrino oscillation experiments, numerous models for neutrino mass have been developed. These models are constructed based on different discrete symmetries such as $S_4$, $A_4$, $\Delta(27)$ and $T_7$ etc. \cite{hirsch2009a4,carcamo2021controlled,sruthilaya2018a_4,borah2019linear,T71,T72,T73,T74,T75,T76,T77,T78,T79,T710,T713,R1,R2,R3,R4,R5,R6,R7,R8,R9,R10,R11,R12}. In most cases, models extending the Standard Model (SM) incorporate symmetries through the addition of specific field contents with their respective charges. These models predict distinctive neutrino masses and mixing patterns by employing different seesaw mechanisms (Type I, II, and III). However, the current model introduces a new aspect by using a hybrid seesaw mechanism combining Type I and Type II, which is not commonly explored in earlier studies focusing on $A_4$ symmetry. This hybrid approach offers a novel pathway to derive neutrino masses and mixing parameters, contributing to the uniqueness of the model compared to previous work.

Despite this innovation, deriving analytical expressions in such models remains challenging due to the complexity of the equations involved. These equations often contain nonlinear terms relating to the neutrino mass eigenstates, making them difficult to solve without advanced mathematical techniques. Compared to earlier $A_4$ models, the current model tackles these challenges with computational methods, particularly using particle swarm optimization (PSO), which proves to be highly effective for solving such complex optimization problems.
The collective actions of fish and birds serve as the model for this population-based algorithm and metaheuristic approach. PSO is used for approximating parameters in different types of research problems \cite{toushmalani2013gravity,bassi2011automatic}. In 1995, Russell Eberhart and James Kennedy introduced the concept of PSO. \cite{panda2008comparison,okwu2021particle,yu2020comparison}, drawing inspiration from genetic algorithms (GAs) to refine its design \cite{panda2008comparison}. PSO is commonly used to find optimal solutions to optimization problems, where the aim is to minimize or maximize a particular fitness function. PSO is versatile and has been utilized in diverse optimization scenarios, including engineering design \cite{guedria2016improved, kaveh2013engineering,kumar2021design}, image processing \cite{djemame2019solving,singh2014image,pramanik2015image},  financial modeling \cite{azayite2019financial, chiam2009memetic,pan2022design, marinakis2009ant} and neural network training \cite{meissner2006optimized, rauf2018training}. Its efficacy is further underscored by its widespread use in diverse optimization challenges, encompassing high-dimensional data clustering \cite{esmin2015review,rana2011review}, parameter estimation for chaotic maps \cite{he2007parameter,alatas2009chaos}, optimization of core loading models in nuclear reactors \cite{babazadeh2009optimization}, optimization of nonlinear reference frames \cite{ibrahim2019hybridization}, attainment of optimal reactive power distribution \cite{subbaraj2010hybrid}, as well as problem-solving in domains such as optical properties of multilayer thin films \cite{jiang2020multilayer,yue2019determination,rabady2014global,ruan2016determination} and autoregressive models with moving average \cite{mehmood2019nature,yetis2014forecasting,yassin2016binary}. Additionally, PSO has proven effective in addressing challenges related to parameter estimation in electromagnetic plane waves \cite{akbar2019novel}. Its simplicity, ease of implementation, and ability to handle non-linear and complex objective functions make it a popular choice for solving optimization problems. PSO stands out in particular for having an easy-to-implement architecture and requiring less memory \cite{stacey2003particle,eberhart1995new}. 

After calculating the mass eigenvalues and lepton mixing matrix upto second order perturbation theory in the framework \cite{Model2023} based on $A_{4}$ symmetry, this study seeks to investigate the minimization of the scalar potential for VEVs and optimize the parameters for $U_{PMNS}$ matrix, neutrino masses and effective neutrino mass parameters: $\langle m_{ee}\rangle,\ m_{\beta}$, for both mass hierarchy through particle swarm optimization (PSO). The format of this article is as follows: The $A_4$ model is presented in the next section \ref{sec:The Model}. In addition to describe the superpotential terms for charged leptons and neutrinos, subsection \ref{subsec:Charged lepton and neutrino mass matrices} provide the explanation for the 
mass eigenvalues and mixing matrix upto second order perturbation theory. Section \ref{sec:Numericsal Analysis} focuses on the utilization of PSO to determine optimal parameter values for computing neutrino masses. Section \ref{sec:potential analysis} presents the scalar potential invariant under $SU(2)_L \times U(1)_Y \times A_{4}\times Z_{10} \times Z_{3}$, along with conditions for its minimization and explores the utilization of PSO in determining optimal parameter values for VEVs of the scalars with addition of $S_2$ flavour symmetry. At the end, in section \ref{sec:Conclusion}, we provide a conclusion of our research.

\section{The \texorpdfstring{$A_4$}{TEXT} based Model }
\label{sec:The Model}
In reference \cite{Model2023}, the authors extended the SM group with $A_4$
symmetry with three 
right handed heavy singlet neutrino 
fields $(\nu_{e_R} ,\nu_{\mu_R} 
,\nu_{\tau_R})$ and with seven scalars 
$\phi$, $\Phi$, $\eta$, $\kappa$, $\Delta$, 
$\xi$, $\xi^{\prime}$. The $SU(2)_L$ doublets $\phi$, $\Phi$ and $SU(2)_L$ triplet $\Delta$ are taken as $A_4$ triplet. Four $SU(2)_L$ singlets $\eta$, $\kappa$, 
$\xi$ and $\xi^{\prime}$ are taken as the singlets of $A_4$ as $1^{\prime\prime}$, $1^{\prime}$, $1$ and $1$ respectively. A additional symmetry, namely $Z_{10}$ is also 
introduced to incorporate the 
undesired terms, where, $Z_{10}$ 
refers to the symmetry of integers 
modulo 10. A summary of all the 
fields under $SU(2)_L$, $A_4$, $Z_{10}$, $Z_{3}$ 
and $U(1)_Y$ are shown in 
table \ref{tab:charge assignments}.
\begin{table}[htbp]
\centering
\begin{tabular}{lllllllllll}
\hline
 Fields& $D_{l_{L}}$ & $l_R$ &$\nu_{l_{R}}$  &$\phi$  & $\Phi$ & $\eta$ & $\kappa$ &  $\Delta$ &$\xi$&  $\xi^{\prime}$ 
 \\
\hline
$SU(2)_L$ & $2$ & $1$&  $1$&  $2$&$2$& 1 & $1$& $3$ &$1$ & $1$\\
\hline
$A_4$ & $3$ & $(1, 1^{\prime\prime}, 1^{\prime})$&  $(1, 1^{\prime\prime}, 1^{\prime})$&  $3$&$3$& $1^{\prime\prime}$ & $1^{\prime}$&$3$ &$1$ & $1$\\
\hline
$U(1)_Y$ & $-1$ & $-2$&  $0$&  $1$&$-1$& $0$ & $0$& $-2$&$0$ & $0$\\
\hline
$Z_{10}$ & $0$ & $0$&  $(0, 4, 6)$& $0$&$0$& $2$ & $8$& $0$&$6$ & $4$\\
\hline
$Z_{3}$ & $1$ & $(\omega,\omega,\omega)$&  $(1, 1, 1)$& $\omega^2$&$1$& $1$ & $1$& $1$&$1$ & $1$\\
\hline
\end{tabular}
\caption{The properties of transformation under $SU(2)_L \times U(1)_Y \times A_{4}\times Z_{10} \times Z_{3}$.\label{tab:charge assignments}}
\end{table}
we also incorporate an additional $S_2$ symmetry. The $S_2$ group, known as the permutation group of two elements, has specific generators when expressed in the 3-dimensional representation, which are as follows:
\[
e = \begin{pmatrix} 
1 & 0 & 0 \\ 
0 & 1 & 0 \\ 
0 & 0 & 1 
\end{pmatrix}, \quad
a = \begin{pmatrix} 
1 & 0 & 0 \\ 
0 & 0 & 1 \\ 
0 & 1 & 0 
\end{pmatrix}
\] 
By choosing the following couplings and imposing $S_2$
flavor symmetry, the resulting constraints are as follows:
\[
y_2 \leftrightarrow y_3,\quad y_{R_1} \leftrightarrow y_{R_2}, \quad \xi \leftrightarrow \xi', \quad \eta \leftrightarrow \kappa
\]
All remaining fields transform trivially under $S_2$.
\subsection{Mass matrices of charged lepton and neutrino}
\label{subsec:Charged lepton and neutrino mass matrices}
\begin{figure}
    \centering
\includegraphics[width=0.48\linewidth]{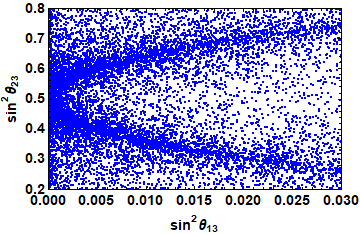} \quad \includegraphics[width=0.48\linewidth]{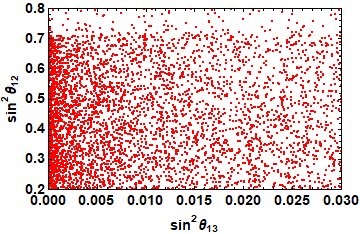}\\ 
\includegraphics[width=0.48\linewidth]{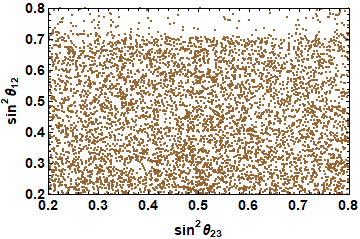} \quad \includegraphics[width=0.48\linewidth]{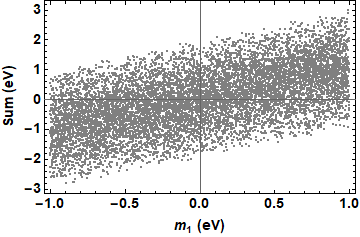} \\
\includegraphics[width=0.62\linewidth]{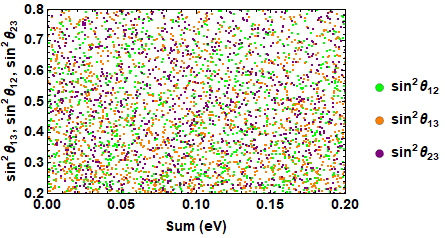}
\caption{The correlation plots include $sin^2\theta_{13}$ versus $sin^2\theta_{23}$ and $sin^2\theta_{12}$, $sin^2\theta_{23}$ versus $sin^2\theta_{12}$, $m_1$ (neutrino mass in $eV$ without perturbations) versus the sum of perturbed neutrino masses, and the sum of perturbed neutrino masses (in $eV$) versus $sin^2\theta_{13}$, $sin^2\theta_{12}$, and $sin^2\theta_{23}$ as shown in last plot.}
\label{fig:Correlations}
\end{figure}

The Lagrangian serves as a cornerstone in describing the interactions and behaviors of particles within the context of physics of particles. The superpotential term for charged leptons, Dirac neutrinos and right handed Majorana neutrinos is given as 
\begin{equation}
\label{eq:superptentialterm}
\begin{aligned}
  - \mathcal{L}_Y &= y_e (\bar{D}_{l_L}\phi)_{1}\,e_{R} + y_{\mu} (\bar{D}_{l_L}\phi)_{1^{\prime}} \,\mu_{R}+\,y_{\tau} (\bar{D}_{l_L}\phi)_{1^{\prime\prime}} \tau_{R}+ y_1\,(\bar{D}_{l_L}\Phi)_1\,\nu_{e_R} \,\\&+ \frac{y_2}{\Lambda}\,(\bar{D}_{l_L}\Phi)_{1^{\prime}} \,\nu_{{\mu}_{R}}\,\xi + \frac{y_3}{\Lambda}\,(\bar{D}_{l_L}\Phi)_{1^{\prime\prime}}\nu_{{\tau}_{R}}\,\xi^{\prime} +\,\frac{1}{2}\,M_1\,({\nu^c_{e_R}}\,\nu_{e_R}) +\,\frac{1}{2}\,M_2\,[({\nu^c_{\mu_{R}}}\nu_{\tau_{R}})\\& \,+ ({\nu^c_{\tau_{R}}} \nu_{\mu_{R}})]+\,\frac{1}{2}\,y_{R_1}({\nu^c_{\mu_{R}}}\,\nu_{\mu_{R}})\,\eta +\frac{1}{2}\,y_{R_2}({\nu^c_{\tau_R}}\nu_{\tau_{R}})\,\kappa +\, y(\bar{D}_{l_L}\, D_{l_L}^c)\,{i \sigma_2 \Delta} \,+ h.c.,
\end{aligned}
\end{equation}
In this context, $y_e$, $y_{\mu}$, $y_{\tau}$, $y_{1,2,3}$, $M_{1,2}$, $y_{R_{1,2}}$ and $y$ are represent the dimensionless Yukawa couplings. Due to the VEVs (see section \ref{sec:potential analysis}) and $S_2$ flavour symmetry, one can generate mass matrices for charged leptons ($M_l$), Dirac neutrinos ($M_D$) and right handed Majorana neutrinos as
\begin{equation}
\label{eq:massmatrices}
\begin{aligned}
&M_{l}=v 
  \begin{pmatrix}
    y_{e} & 0 & 0\\
       0 & y_{\mu} & 0\\
       0 & 0 & y_{\tau}\\
  \end{pmatrix}
\end{aligned}, \qquad \begin{aligned}
M_{D}=u 
  \begin{pmatrix}
    0 & \frac{y_{2}v_\epsilon}{\Lambda} & \frac{y_{2}v_\epsilon}{\Lambda}\\
       y_{1} & 0 & \frac{y_{2}v_\epsilon}{\Lambda}\\
       y_{1} & \frac{y_{2}v_\epsilon}{\Lambda} & 0\\
  \end{pmatrix}
\end{aligned},
\end{equation}
\begin{equation}
\label{eq:masmatrices}
\begin{aligned}
M_{R}=
  \begin{pmatrix}
    M_1 & 0 & 0\\
       0 & y_Rv_m & M_2\\
       0 & M_2 & y_Rv_m\\
  \end{pmatrix}, \end{aligned} \qquad \begin{aligned}
M^{\prime\prime}=
 \frac{y\omega}{3} \begin{pmatrix}
    0 & 1 & -1\\
       1 & 2 & 0\\
       -1 & 0 & -2\\
  \end{pmatrix}.
\end{aligned}
\end{equation}
Here, seesaw frameworks, particularly type-I \cite{cai2018lepton, mohapatra2005seesaw, king2013neutrino, mohapatra2006neutrino, king2003neutrino} and type-II \cite{melfo2012type, perez2008neutrino, cheng1980neutrino}, used besides several other methods to explain small neutrino masses. Majorana and Dirac mass terms derived from the introduction of extra right-handed neutrinos in SM in type I. Majorana mass terms ($M^{\prime\prime}$) derived from the introduction of heavy $SU(2)_L$ triplet in SM in type-II. In other words, a hybrid seesaw mechanism \cite{cai2018lepton, akhmedov2007interplay, wong2022tree} proposed for improved mass suppression and new mixing patterns by combining type-I and type-II. With this hybrid technique, one may explore various lepton mixing scenarios and generate effective Majorana neutrino mass matrices $(M_\nu)$ as
\begin{equation}
\begin{aligned}
\label{eq:neutrinomasses}
M_\nu=M^{\prime}+M^{\prime\prime}=-m_D M_R^{-1} m_D^{T}+M^{\prime\prime}
\end{aligned}
\end{equation}
with
\begin{equation}
\label{eq:A}
\begin{aligned}
M^{\prime}&=
  \begin{pmatrix}
    P & Q & Q\\
       Q& R & S\\
      Q & S & R
  \end{pmatrix}, \qquad M^{\prime\prime}=\begin{pmatrix}
    0 & p & -p\\
       p & q & 0\\
       -p & 0 & -q\\
  \end{pmatrix},\\
p&=\frac{y\omega}{3},\quad q=2p,\quad
P=-\frac{2 u^2 {v_\epsilon}^2 {y_2}^2}{\Lambda^2 (M_2+v_m y_R)}, \quad Q=-\frac{u^2 {v_\epsilon}^2 {y_2}^2}{\Lambda ^2 (M_2+v_m y_R)},\quad
\\ R&=u^2 \left(\frac{v_m {v_\epsilon}^2 {y_2}^2 y_R}{\Lambda ^2 \left({M_2}^2-{v_m}^2 {y_R}^2\right)}-\frac{{y_1}^2}{{M_1}}\right), \quad S=u^2 \left(-\frac{{y_1}^2}{{M_1}}-\frac{{M_2} {v_\epsilon}^2 {y_2}^2}{\Lambda ^2 \left({M_2}^2-{v_m}^2 {y_R}^2\right)}\right).
\end{aligned}
\end{equation}
The first effective Majorana matrix is diagonalized under the conditions $y_2=y_3$, $y_{R_1}=y_{R_2}$ and $S_2$ symmetry, by the subsequent mixing matrix,
\begin{equation}
\label{eq:V}
\begin{aligned}
U_0=
  \begin{pmatrix}
    c & s & 0\\
       -s/\sqrt{2}& c/\sqrt{2} & 1/\sqrt{2}\\
      -s/\sqrt{2} & c/\sqrt{2} & -1/\sqrt{2}
  \end{pmatrix},
\end{aligned}
\end{equation}
such as $diag(m_1,\ m_2,\ m_3)={U_0}^T M_{1}\ {U_0}$, where, $c=\cos\theta$,
$s=\sin\theta$ and $\theta=arccos(\frac{k}{\sqrt{k^2+2}})$ with, 
\begin{equation}
\label{eq:k}
\begin{aligned}
k=\frac{P-R-S-\sqrt{P^2-2 P R-2 P S+8 Q^2+R^2+2 R S+S^2}}{2 Q}
\end{aligned},
\end{equation}
and
\begin{equation}
\label{eq:ma}
\begin{aligned}
m_{1,2}=\frac{1}{2} (P+R+S\mp\sqrt{(-P+R+S)^2+8 Q^2}), \qquad m_3=R-S.
\end{aligned}
\end{equation}
In the context of three-neutrino physics, the mixing matrix of lepton $(U_{PMNS})$ may be represented as \cite{vien2016delta}
\begin{equation}
\label{eq:para}
\begin{aligned}
U_{PMNS}= 
  \begin{pmatrix}
    c_{12}c_{13} & s_{12}c_{13} & s_{13} e^{i\delta}\\
    -s_{12}c_{23}-c_{12}s_{13}s_{23}e^{i\delta} & c_{12}c_{23}-s_{12}s_{13}s_{23}e^{i\delta} & c_{13}s_{23}\\
    s_{12}s_{23}-c_{12}c_{23}s_{13}e^{i\delta} & -c_{12}s_{23}-s_{12}s_{13}c_{23}e^{i\delta} & c_{13}c_{23}
  \end{pmatrix} P_{12}
\end{aligned}
\end{equation} 
where, $P_{12}=diag(1, e^{i\beta_{1}}, e^{i\beta_{2}})$
which contains two Majorana phases that do not influence neutrino oscillations. 
The matrix $U_0$ in equation \ref{eq:V} suggests $\theta_{23} = \pi/4$, $\theta_{13} = 0$ and $\theta_{12} = \theta$, Recent data contradicts this claim. However, the inclusion of the second matrix in equation \ref{eq:neutrinomasses} is expected to ameliorate this discrepancy. In first-order perturbation corrections, the second matrix in equation \ref{eq:neutrinomasses} doesn't affect the eigenvalues but it does influence the eigenvectors. Moving to second-order perturbation theory, this matrix contributes to the determination of both eigenvalues and eigenvectors. Consequently, the masses of neutrino upto the second order perturbation corrections can be expressed as:
\begin{equation}
\label{eq:2ndordercorrection}
\begin{aligned}
m^{\prime}_1=m_1+\frac{p^2{\Gamma_1}^2}{m_1-m_3},\quad
m^{\prime}_2=m_2+\frac{p^2{\Gamma_2}^2}{m_2-m_3},\quad
m^{\prime}_3=m_3+p^2\left(\frac{2{\Gamma_3}^2}{m_3-m_1}+\frac{{\Gamma_2}^2}{m_3-m_2}\right),
\end{aligned}
\end{equation}
where the parameters $p$, $q$, and $m_{1,2,3}$ are defined in equations \ref{eq:A} and \ref{eq:ma}, respectively.
The contributions to the eigenvalues from first-order perturbation are negligible, with only minor changes arising at second-order perturbation. In Eq. (2.10), the terms $\frac{p^2{\Gamma_1}^2}{m_1-m_3}$, $\frac{p^2{\Gamma_2}^2}{m_2-m_3}$, and $p^2\left(\frac{2{\Gamma_3}^2}{m_3-m_1}+\frac{{\Gamma_2}^2}{m_3-m_2}\right)$ are significantly smaller compared to $m_1$, $m_2$, and $m_3$, respectively. This is consistent with the condition $M^{\prime\prime} \ll M^{\prime}$.  
Subsequently, the resulting lepton mixing matrix is as follows:
\begin{equation}
\label{eq:2ndordereigenvector}
\begin{aligned}
U=U_0+\Delta U+\Delta U^{\prime} 
\end{aligned}
\end{equation}
where $U_0$ is given by  in equation \ref{eq:V}, $\Delta U$ represents the mixing matrix corresponding to first-order corrections, and $\Delta U^{\prime}$ represents the mixing matrix corresponding to second-order corrections. They have the following entries:
\begin{equation}
\label{eq:entries1}
\begin{aligned}
(\Delta U)_{11}&=(\Delta U)_{12}=0,
\end{aligned}\nonumber
\end{equation}
\begin{equation}
\label{eq:entries2}
\begin{aligned}
(\Delta U)_{13}&=p\left(\frac{\Gamma_1 \cos{\theta}}{{m_3}-{m_1}}+\frac{\Gamma_2\sin{\theta}}{{m_3}-{m_2}}\right),
\end{aligned}\nonumber
\end{equation}
\begin{equation}
\label{eq:entries3}
\begin{aligned}
(\Delta U)_{21}=-(\Delta U)_{31}=\frac{p\Gamma_3}{{m_3}-{m_1}},\quad (\Delta U)_{32}=-(\Delta U)_{22}=\frac{\sqrt{2}\Gamma_2 p}{2({m_3}-{m_2})},
\end{aligned}\nonumber
\end{equation}
\begin{equation}
\label{eq:entries5}
\begin{aligned}
(\Delta U)_{23}&=(\Delta U)_{33}=\frac{p \left(({m_1}-{m_2})\Gamma_4+2 ({m_1}+{m_2}-2 {m_3})\right)}{2 \sqrt{2} ({m_3}-{m_1}) ({m_2}-{m_3})},
\end{aligned}\nonumber
\end{equation}
\begin{equation}
\begin{aligned}
\label{eq:entriesprime1}
(\Delta U^{\prime})_{11}=\frac{p^2\Gamma_1}{2 ({m_1}-{m_3})^2} \biggl[-\cos \theta\Gamma_1+\frac{2 \sin \theta ({m_1}-{m_3})\Gamma_2}{{m_1}-{m_2}}\biggl],
\end{aligned}\nonumber
\end{equation}
\begin{equation}
\begin{aligned}
\label{eq:entriesprime2}
(\Delta U^{\prime})_{12}&=\frac{p^2\Gamma_2}{2 ({m_2}-{m_3})^2} &\biggl[-\sin \theta \Gamma_2 -\frac{2 \cos \theta ({m_2}-{m_3})\Gamma_1}{{m_1}-{m_2}}\biggl], \quad (\delta U^{\prime})_{13}=0,
\end{aligned}\nonumber
\end{equation}
\begin{equation}
\begin{aligned}
\label{eq:entriesprime3}
(\Delta U^{\prime})_{21}=(\Delta U^{\prime})_{31}=p^2 \Gamma_1&\biggl[\frac{\Gamma_5 (3 {m_1}-{m_2}-2 {m_3})+{m_1}+{m_2}-2 {m_3})}{{2 \sqrt{2} ({m_1}-{m_2}) ({m_1}-{m_3})^2} }\biggl],
\end{aligned}\nonumber
\end{equation}
\begin{equation}
\begin{aligned}
\label{eq:entriesprime4}
(\Delta U^{\prime})_{22}=(\Delta U^{\prime})_{32}=-p^2 \Gamma_2\biggl[\frac{\Gamma_5 ({m_1}-3 {m_2}+2 {m_3})+{m_1}+{m_2}-2 {m_3}}{2 \sqrt{2} ({m_1}-{m_2}) ({m_2}-{m_3})^2}\biggl],
\end{aligned}\nonumber
\end{equation}
\begin{equation}
\begin{aligned}
\label{eq:entriesprime5}
(\Delta U^{\prime})_{23}=&-(\Delta U^{\prime})_{33}=\frac{-p^2}{2 \sqrt{2}}\biggl[\frac{{\Gamma_1}^2}{({m_1}-{m_3})^2}+\frac{{\Gamma_2}^2}{({m_2}-{m_3})^2}\biggl],
\end{aligned}\nonumber
\end{equation}
with, $\Gamma_1=-2\sin{\theta}+\sqrt{2}\cos{\theta}$, $\Gamma_2=\sqrt{2}\sin{\theta}+2\cos{\theta}$, $\Gamma_3=\sqrt{2}\sin{\theta}-\cos{\theta}$, $\Gamma_4=\sqrt{2}\sin{2\theta}+2\cos{2\theta}$ and $\Gamma_5=\cos{2\theta}+\sqrt{2}\sin{\theta}\cos{\theta}$.
The lepton mixing angles can be determined from equations \ref{eq:2ndordereigenvector} and \ref{eq:para}, which define the mixing matrix of  neutrino:
\begin{equation}
\label{eq:for}
\begin{aligned}
t_{12}=\frac{|U_{12}|}{|U_{11}|},\quad t_{23}=\frac{|U_{23}|}{|U_{33}|}, \quad s_{13}=|U_{13}|,
\end{aligned}
\end{equation}
with, $s_{ij}=\sin{\theta_{ij}}$, $c_{ij}=\cos{\theta_{ij}}$ and $t_{ij}=\tan{\theta_{ij}}$. Most of $A_4$ models predict $\theta_{13}=0$. In this model, there is a deviation, for sure there is some correlation with other observable like $\theta_{12}$, $\theta_{23}$. The correlations between different parameters are given in figure \ref{fig:Correlations}.
\section{Numerical Analysis}
\label{sec:Numericsal Analysis}
Taking into consideration the latest experimental data \cite{king2013neutrino}, the mixing angles are measured as follows: The solar neutrino mixing angle, $\theta_{12}$, is determined to be $34^\circ \pm 1^\circ$, the atmospheric neutrino mixing angle, $\theta_{23}$, is found to be $42^\circ \pm 3^\circ$, and the reactor angle, $\theta_{13}$, is measured to be $8.5^\circ\pm 0.5^\circ$. Additionally, the squared mass differences are determined as $\Delta m_{\text{sol}}^2 = m_2^{{\prime}^2} - m_1^{{\prime}^2} \approx 7.53 \times 10^{-5}$ eV$^2$ and $\Delta m_{\text{atm}}^2 = m_3^{{\prime}^2} - m_2^{{\prime}^2} \approx 2.453 \times 10^{-3}$ eV$^2$ ($\Delta m_{\text{atm}}^2 = m_3^{{\prime}^2} - m_2^{{\prime}^2} \approx -2.536 \times 10^{-3}$ eV$^2$) for normal (inverted) neutrino mass ordering  \cite{particle2022review}. The lower and upper bounds of $\Sigma m^{\prime}$ are constrained to $0.06$ eV and $0.12$ eV, respectively \cite{katrin2022direct}. Utilizing equations \ref{eq:2ndordereigenvector} and \ref{eq:for}, the objective or fitness function $(\epsilon)$ corresponding to these experimental constraints can be expressed as follows.
\begin{equation}
\label{eq:sum}
\begin{aligned}
\epsilon=\epsilon_1+\epsilon_2+\epsilon_3+\epsilon_4+\epsilon_5+\epsilon_6,
\end{aligned}
\end{equation}
with,
\begin{equation}
\label{eq:objectivefunctions}
\begin{aligned}
\epsilon_{1}&=\Biggl[{m^{\prime}_2}^2-{m^{\prime}}_1^2-\text{$\Delta $m}_{\text{sol}}^2\Biggl]^2,\quad 
\epsilon_{2}=\Biggl[{m^{\prime}_3}^2-{m^{\prime}}_2^2-\text{$\Delta $m}_{\text{atm}}^2\Biggl]^2,\quad
\epsilon_{3}=\Biggl[\frac{|U_{12}|}{|U_{11}|}-t_{12}\Biggl]^2,\\
\epsilon_{4}&=\Biggl[\frac{|U_{23}|}{|U_{33}|}-t_{23}\Biggl]^2,\quad
\epsilon_{5}=\Biggl[U_{13}-s_{13}\Biggl]^2,\\
\epsilon_{6}&=\Biggl[m^{\prime}_1+m^{\prime}_2+m^{\prime}_3-\Biggl(\begin{matrix}
0.12\ eV,\quad \text{for upper bound limit} \\
0.06\ eV,\quad  \text{for lower bound limit} \\ 
\end{matrix}\Biggl)\Biggl]^2,
\end{aligned}
\end{equation}
where, $m^{\prime}_{1,2,3}$, $U_{11}$, $U_{12}$, $U_{23}$, $U_{33}$, $U_{13}$ are given in equations \ref{eq:2ndordercorrection} and \ref{eq:2ndordereigenvector}.

In this article we use PSO instead of classical Chi-square fitting for optimization and parameter estimation tasks because of following reasons: 1)  PSO is a population-based optimization technique that searches the entire solution space for the optimal solution. It can escape local optima more effectively compared to Chi-square fitting, which might get stuck in local minima. 2) PSO tends to converge faster than Chi-square fitting. 3) PSO does not require initial guesses for the parameters being optimized, unlike Chi-square fitting, where a good initial guess is required for convergence to the proper solution. 4) PSO can handle non-linear optimization problems efficiently, which may be challenging for Chi-square fitting. 5) PSO is highly adaptable and can be easily modified or extended to handle various types of optimization tasks, including parameter estimation, feature selection, and function optimization. 6) PSO can be parallelized effectively, allowing for faster computation times on parallel computing architectures compared to Chi-square fitting, which may not be as easily parallelized. 7) PSO is robust to noise and can handle noisy objective functions better than some Chi-square fitting methods, which may struggle with noisy data.

We use PSO algorithm to optimize the objective or fitness functions $\epsilon$ for optimal values of parameters. In the usual way of doing particle swarm optimization (PSO), we treat each possible solution to a problem is represented as a moving point within the search space. These points form a group called a 'swarm', and they work together to check out the whole search area. Each point is assigned a unique score based on its efficacy in solving the problem. Initially, these points are randomly selected. Throughout each iteration, the positions and velocities of the points are updated based on their previous performance based on its local $P^{x-1}_{LB}$ and global $P^{x-1}_{GB}$ positions. The basic rules for updating position and velocity of a point are given as,
\begin{equation}
\label{eq:PSO1}
\begin{aligned}
v^{x}_i=wv^{t-1}_i+c_1 r_1(P^{x-1}_{LB}-X^{x-1}_{i})+c_2 r_2(P^{x-1}_{GB}-X^{x-1}_{i}),
\end{aligned}
\end{equation}
\begin{equation}
\label{eq:PSO2}
\begin{aligned}
X^{x}_i=X^{x-1}_i+v^{x-1}_i.
\end{aligned}
\end{equation}
In these rules, $i$ goes from $1$ to $p$, where $p$ is just a integer telling us how many points there are.  The weight '$w$' and $c_1$ and $c_2$ are also integers that help to control how the points move. Also, the velocity gradually gets smaller as we keep looking around (between 0 and 1). The random numbers $r_1$ and $r_2$ are just there to add a bit of randomness. Finally, the velocity of the points is kept within certain limits so they don't go too fast or too slow.

The points traverse the search space by adapting their positions and velocities, drawing from their individual experiences and insights gained from neighboring points. The algorithm involves the following key steps:
\begin{enumerate}
\item Initialization: Commence by populating a set of points, assigning them random positions and velocities distributed throughout the exploration area.
\item Objective Assessment: Assess the fitness or objective function value for each point according to its present position.
\item Update Personal and Global Bests: Update the personal best position (Pbest) for each point based on its current fitness. Update the global best position (Gbest) considering the best position among all points.
\item Update Velocities and Positions: Adjust the velocity and position of each point using its current velocity, personal best, and global best positions.
\item Iteration: Continue steps 2 through 4 for a predetermined number of iterations or until reaching a convergence criterion.

The generic flow chart PSO is given in figure \ref{fig:PSO1}.
\begin{figure}
    \centering
    \includegraphics[width=0.9\linewidth]{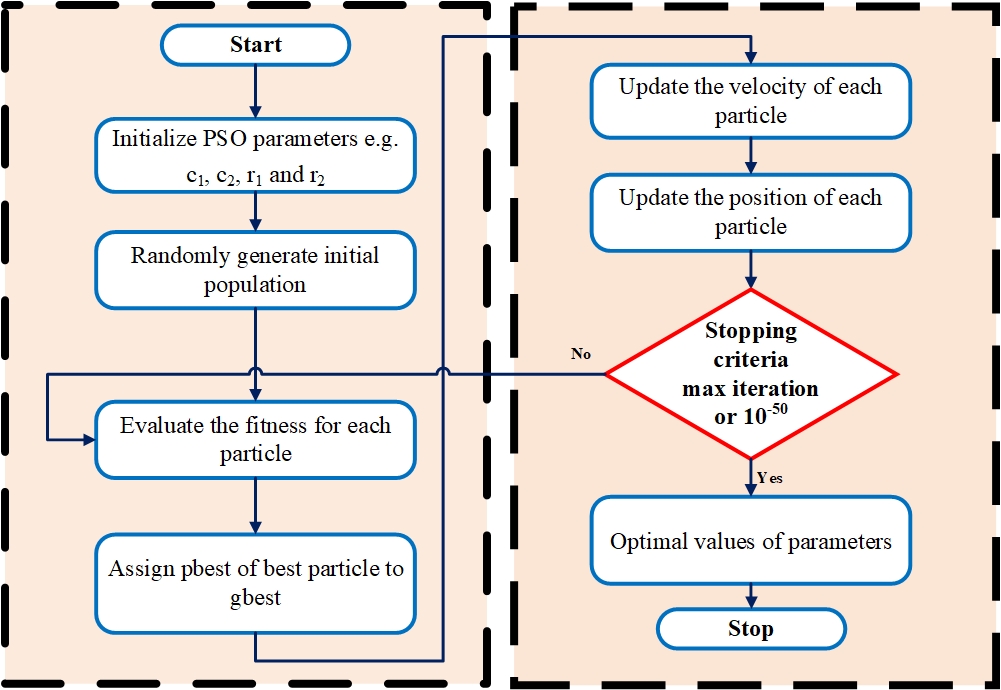}
    \caption{Generic flow chart of PSO}
    \label{fig:PSO1}
\end{figure}
\end{enumerate}

\begin{figure}
    \centering
    \includegraphics[width=1\linewidth]{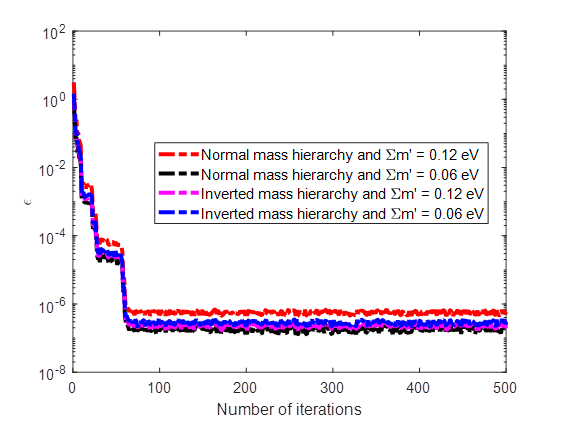}
    \caption{Fitness function versus number of iterations}
    \label{fig:Fitness function versus number of iterations}
\end{figure}

To inspire the development of meta-heuristic optimization algorithms, we employed PSO technique to minimize the objective function for both mass hierarchy and for upper and lower bound limits of $\Sigma m^{\prime}$. The objective function is minimized through PSO with 500 iteration are presented in figure \ref{fig:Fitness function versus number of iterations} and corresponding values of $p$, $\theta$, $m_1$, $m_2$ and $m_3$ are given in table \ref{tab:optimal values for Normal hierarchy and upper bound limit of}, \ref{tab:optimal values for Normal hierarchy and lower bound limit of}, \ref{tab:optimal values for Inverted hierarchy and upper bound limit of} and \ref{tab:optimal values for Inverted hierarchy and lower bound limit of} in which $m_1$, $m_2$ and $m_3$ are masses of neutrinos in $eV$ without perturbations correspond to zero reactor angle and $p$ and $\theta$ are the free parameters of the model. The masses of neutrino without perturbations ($m_1$, $m_2$, $m_3$) and other model free parameters $p$ and $\theta$ are predicted using PSO by using experimental squared mass differences of perturbed masses.
\begin{table}[htbp]
\centering
\begin{tabular}{llll}
\hline
 Parameters& Optimal values & Parameters &Optimal values
 \\
\hline
\hline
$p$ & $-4.5205\times10^{-16}$ & $\theta$&  $0.5894$ $rad$\\
\hline
$m_1$ & $4.0000\times 10^{-2}\ eV$&  $m_2$& $4.0000\times 10^{-2}\ eV$\\
\hline
$m_3$ & $4.0000\times 10^{-2}\ eV$ & & \\
\hline
\end{tabular}
\caption{The optimal values of parameters $p$, $\theta$, $m_1$, $m_2$, $m_3$, through PSO for upper bound limit of $\Sigma m^{\prime}=0.12\ eV$ and normal mass hierarchy.\label{tab:optimal values for Normal hierarchy and upper bound limit of}}
\end{table}
\begin{table}[htbp]
\centering
\begin{tabular}{llll}
\hline
 Parameters& Optimal values & Parameters &Optimal values
 \\
\hline
\hline
$p$ & $1.0964\times10^{-14}$ & $\theta$&  $0.5909$ $rad$\\
\hline
$m_1$ & $2.0000\times 10^{-2}\ eV$&  $m_2$& $2.0000\times 10^{-2}\ eV$\\
\hline
$m_3$ & $2.0000\times 10^{-2}\ eV$ & & \\
\hline
\end{tabular}
\caption{The optimal values of parameters $p$, $\theta$, $m_1$, $m_2$, $m_3$, through PSO for lower bound limit of $\Sigma m^{\prime}=0.06\ eV$ and normal mass hierarchy.\label{tab:optimal values for Normal hierarchy and lower bound limit of}}
\end{table}
\begin{table}[htbp]
\centering
\begin{tabular}{llll}
\hline
 Parameters& Optimal values & Parameters &Optimal values
 \\
\hline
\hline
$p$ & $1.0000\times 10^{-4}$ & $\theta$&  $0.5909$ $rad$\\
\hline
$m_1$ & $3.8627 \times 10^{-2}\ eV$&  $m_2$& $4.0517\times 10^{-2}\ eV$\\
\hline
$m_3$ & $3.8563\times 10^{-2}\ eV$ & & \\
\hline
\end{tabular}
\caption{The optimal values of parameters $p$, $\theta$, $m_1$, $m_2$, $m_3$, through PSO for inverted mass hierarchy and for upper bound limit of $\Sigma m^{\prime}=0.12\ eV$.\label{tab:optimal values for Inverted hierarchy and upper bound limit of}}
\end{table}
\begin{table}[htbp]
\centering
\begin{tabular}{llll}
\hline
 Parameters& Optimal values & Parameters &Optimal values
 \\
\hline
\hline
$p$ & $1.4714\times 10^{-3}$ & $\theta$&  $0.5909$ $rad$\\
\hline
$m_1$ & $1.1040\times 10^{-2}\ eV$&  $m_2$& $3.8846\times 10^{-2}\ eV$\\
\hline
$m_3$ & $1.0098\times 10^{-2}\ eV$ & & \\
\hline
\end{tabular}
\caption{The optimal values of parameters $p$, $\theta$, $m_1$, $m_2$, $m_3$, through PSO for lower bound limit of $\Sigma m^{\prime}=0.06\ eV$ and inverted mass hierarchy.
\label{tab:optimal values for Inverted hierarchy and lower bound limit of}}
\end{table}

The lepton mixing matrices upto second order perturbation theory for both mass hierarchy and for upper ($0.12\ eV$) and lower ($0.06\ eV$) bound limits of $\Sigma m^{\prime}$ are given as:
\begin{equation}
\label{eq:UPMNS Normal and upper bound}
\begin{aligned}
|U^{(N)_{upper}}_{PMNS}|=\left(
\begin{array}{ccc}
 0.8208 & 0.5519 & 0.1476 \\
 0.3317 & 0.6710 & 0.6617 \\
 0.4653 & 0.4952 & 0.7353 \\
\end{array}
\right)
\end{aligned}
\end{equation}
\begin{equation}
\label{eq:UPMNS Normal and lower bound}
\begin{aligned}
|U^{(N)_{lower}}_{PMNS}|=\left(
\begin{array}{ccc}
 0.8201 & 0.5539 & 0.1470 \\
0.3336 & 0.6710 & 0.6615 \\
 0.4650 & 0.4938 & 0.7355 \\
\end{array}
\right)
\end{aligned}
\end{equation}
\begin{equation}
\label{eq:UPMNS inverted and upper bound}
\begin{aligned}
|U^{(I)_{upper}}_{PMNS}|=\left(
\begin{array}{ccc}
 0.8200 & 0.5531 & 0.1477 \\
 0.3330 & 0.6710 & 0.6618 \\
 0.4657 & 0.4938 & 0.7350 \\
\end{array}
\right)
\end{aligned}
\end{equation}
\begin{equation}
\label{eq:UPMNS inverted and lower bound}
\begin{aligned}
|U^{(I)_{lower}}_{PMNS}|=\left(
\begin{array}{ccc}
 0.8200 & 0.5531 & 0.1478 \\
 0.3330 & 0.6710 & 0.6618 \\
 0.4658 & 0.4938 & 0.7350 \\
\end{array}
\right)
\end{aligned}
\end{equation}

On the behalf of the values of $p$, $\theta$, $m_1$, $m_2$ and $m_3$, the mass corrections upto second order perturbation theory for both mass hierarchy and for upper ($0.12\ eV$) and lower ($0.06\ eV$) bound limits of $\Sigma m^{\prime}$ are given as: 
${m_{1}^{\prime}}^{(N)}(upper)=4.0000 \times 10^{-2}\ eV$, ${m_{2}^{\prime}}^{(N)}(upper)=4.0000 \times 10^{-2}\ eV$,  
${m_{3}^{\prime}}^{(N)}(upper)=4.0000 \times 10^{-2}\ eV$, ${m_{1}^{\prime}}^{(I)}(upper)=3.8628\times 10^{-2}\ eV$,
${m_{2}^{\prime}}^{(I)}(upper)=4.0548\times 10^{-2}\ eV$,
${m_{3}^{\prime}}^{(I)}(upper)=3.8532\times 10^{-2}\ eV$,
${m_{1}^{\prime}}^{(N)}(lower)=2.0000 \times 10^{-2}\ eV$, 
${m_{2}^{\prime}}^{(N)}(lower)=2.0000 \times 10^{-2}\ eV$, 
${m_{3}^{\prime}}^{(N)}(lower)=2.0000 \times 10^{-2}\ eV$, 
${m_{1}^{\prime}}^{(I)}(lower)=1.1049\times 10^{-2}\ eV$,
${m_{2}^{\prime}}^{(I)}(lower)=3.9298\times 10^{-2}\ eV$ and
${m_{3}^{\prime}}^{(I)}(lower)=9.6381\times 10^{-3}\ eV$, which are correspond to non zero reactor angle. The Dirac CP-violating phase $\delta$ is measured to be $-\pi/2$. In our numerical analysis, we concentrated on only two specific values for the sum of neutrino masses, instead of a continuous spectrum. We utilized 0.06 eV, representing the lower limit of the sum of masses, and 0.12 eV as the upper limit. In the NO case and the neutrino masses appear nearly degenerate and normal hierarchy could not be adequately reproduced here. In contrast, for the inverted ordering, a more defined hierarchy was observable, making the model consistent with an inverted hierarchy (IO). These findings are consistent with a quasi-degenerate normal hierarchy (QDNH) mass spectrum, characterized by ${m_{1}^{\prime}}^N\lesssim{m_{2}^{\prime}}^N\lesssim{m_{3}^{\prime}}^N$, and an inverted mass hierarchy (IH), where ${m_{3}^{\prime}}^I<{m_{1}^{\prime}}^I<{m_{2}^{\prime}}^I$ for both $\Sigma m^{\prime}=0.06$ and $0.12$.  
\subsection{Effective neutrino mass parameters}
\label{sec:Effective neutrino mass parameters}
The expressions for the effective neutrino masses \cite{rodejohann2011neutrino, mitra2012neutrinoless, bilenky2012neutrinoless, rodejohann2012neutrinolessssw,vergados2012theory} associated with neutrinoless double beta decay ($\langle m_{ee} \rangle$) and beta decay ($m_\beta$) are structured as follows:
\begin{equation}
\label{effective majorana parameter}
\begin{aligned}
m_{\beta} =\sqrt{\sum_{i = 1}^{3}{|U_{\text{ei}}|^{2}m_{i}^{{\prime}^2}}}, \quad \left\langle m_{\text{ee}} \right\rangle = \left| \sum_{i = 1}^{3}{U_{\text{ei}}^{2}m_{i}^{\prime}} \right| 
\end{aligned},
\end{equation}

Considering the leptonic mixing matrix elements $U_{ei}$ with $i$ ranging from 1 to 3, representing the masses $m^{\prime}_i$ of three neutrinos, the effective neutrino masses associated with neutrinoless double beta decay ($\langle m_{ee} \rangle$) and beta decay ($m_\beta$) are computed using the parameters obtained in section \ref{sec:Numericsal Analysis}. This calculation is performed for both mass hierarchy of  neutrino, yielding the following results: 

\begin{equation}
\label{effective majorana parametervalues}
\begin{aligned}
\langle {m_{ee}} \rangle^{N}(upper)&=40.0050 \ meV, \qquad
\langle {m_{\beta}} \rangle^{N}(upper)&=40.0025\ meV,\\
\langle {m_{ee}} \rangle^{I}(upper)&=39.2181\ meV, \qquad
\langle {m_{\beta}} \rangle^{I}(upper)&=39.2257\ meV,\\
\langle {m_{ee}} \rangle^{N}(lower)&=20.0024\ meV, \qquad
\langle {m_{\beta}} \rangle^{N}(lower)&=20.0012\ meV,\\
\langle {m_{ee}} \rangle^{I}(lower)&=19.6608\ meV, \qquad \langle {m_{\beta}} \rangle^{I}(lower)&=23.5908\ meV.
\end{aligned}
\end{equation}

\section{Vacuum alignment studies}
\label{sec:potential analysis}
In particle physics, the dynamics of scalar fields are encapsulated by an invariant scalar potential within the symmetry group. The following equation \ref{eq:x1} describes the invariant scalar potential within the symmetry group $SU(2)_L \times U(1)_Y \times A_{4}\times Z_{10} \times Z_{3}$. It plays a crucial role in understanding spontaneous symmetry breaking and the generation of particle masses. While $A_4$, $Z_{10}$ and $Z_{3}$ are discrete symmetries that add to the rich structure of the potential, the $SU(2)_L$ symmetry describes weak isospin.
\begin{equation}
\label{eq:x1}
\begin{aligned}
V=& V(\phi)+ V(\Phi)+V(\Delta)+V(\eta)+ V(\kappa)+V(\xi)+V({\xi^{\prime}})+V(\phi,\Phi)+V(\phi, \Delta)+V(\phi,\eta)\\&+V(\phi,\kappa)+V(\phi,\xi)+V(\phi,{\xi^{\prime}})+V(\Phi, \Delta)+V(\Phi,\eta)+V(\Phi,\kappa)+V(\Phi,\xi)+V(\Phi,{\xi^{\prime}})\\&+V(\Delta,\eta)+V(\Delta,\kappa)+V(\Delta,\xi)+V(\Delta,{\xi^{\prime}})+V(\eta,\kappa)+V(\eta,\xi)+V(\eta,{\xi^{\prime}})+V(\kappa,\xi)\\&+V(\kappa,{\xi^{\prime}})+V(\xi,{\xi^{\prime}})+V(\phi,\eta,\kappa)+V(\phi,\xi,{\xi^{\prime}})+
V(\Phi,\eta,\kappa)+V(\Phi,\xi,{\xi^{\prime}})+V(\Delta,\eta,\kappa) \\&+V(\Delta,\xi,{\xi^{\prime}})+V(\eta,\kappa,\xi,{\xi^{\prime}}).
\end{aligned}
\end{equation}
Writing the terms explicitly, we have,
\begin{align}
V(\phi)&=-\mu^2_{\phi}(\phi^{\dagger}\phi)+ \lambda^{\phi}_1(\phi^{\dagger}\phi)(\phi^{\dagger}\phi)+\lambda^{\phi}_2(\phi^{\dagger}\phi)_{1'}(\phi^{\dagger}\phi)_{1''}+\lambda^{\phi}_3(\phi^{\dagger}\phi)_{3_s}(\phi^{\dagger}\phi)_{3_s},\nonumber\\
V(\Phi)&= V(\phi \rightarrow \Phi),\nonumber\\
V(\Delta)&=-\mu^2_{\Delta}Tr(\Delta^{\dagger}\Delta)+ \lambda^{\Delta}_1 Tr(\Delta^{\dagger}\Delta)Tr(\Delta^{\dagger}\Delta)+\lambda^{\Delta}_2 Tr(\Delta^{\dagger}\Delta)_{1'} Tr(\Delta^{\dagger}\Delta)_{1''}\nonumber\\&\quad\,+\lambda^{\Delta}_3 Tr(\Delta^{\dagger}\Delta)_{3_s} Tr(\Delta^{\dagger}\Delta)_{3_s}\nonumber\\
V(\eta)&=  -\mu^2_{\eta}(\eta^{*}\eta)+ \lambda^{\eta}(\eta^{*}\eta)^2,\quad
V(\kappa)=  -\mu^2_{\kappa}(\kappa^{*}\kappa)+ \lambda^{\kappa}(\kappa^{*}\kappa)^2,\nonumber\\
V(\xi)&=  -\mu^2_{\xi}(\xi^{*}\xi)+ \lambda^{\xi}(\xi^{*}\xi)^2,\quad
V({\xi^{\prime}})=  -\mu^2_{{\xi^{\prime}}}({\xi^{\prime}}^{*}{\xi^{\prime}})+ \lambda^{{\xi^{\prime}}}({\xi^{\prime}}^{*}{\xi^{\prime}})^2,\nonumber\\
V(\phi,\Phi)&=\lambda^{\phi \Phi}_1(\phi^{\dagger}\phi)(\Phi^{\dagger}\Phi)+\lambda^{\phi \Phi}_2((\phi^{\dagger}\phi)_{1'}(\Phi^{\dagger}\Phi)_{1''}+(\phi^{\dagger}\phi)_{1''}(\Phi^{\dagger}\Phi)_{1'})\nonumber\\&+\lambda^{\phi \Phi}_3(\phi^{\dagger}\phi)_{3_s}(\Phi^{\dagger}\Phi)_{3_s},\nonumber\\
V(\phi, \Delta)&= V(\phi, \Phi \rightarrow \phi,\Delta),\quad
V(\phi,\eta)=\lambda^{\phi \eta}(\phi^{\dagger}\phi)\eta^{*}\eta,
 \quad V(\phi,\kappa)=\lambda^{\phi \kappa}(\phi^{\dagger}\phi)\kappa^{*}\kappa,\nonumber\\
V(\phi,\xi)&=\lambda^{\phi \xi}(\phi^{\dagger}\phi)\xi^{*}\xi,\quad
 V(\phi,{\xi^{\prime}})=\lambda^{\phi {\xi^{\prime}}}(\phi^{\dagger}\phi){\xi^{\prime}}^{*}{\xi^{\prime}},\quad
V(\Phi,\Delta)=V(\phi,\Phi \rightarrow \Phi,\Delta),\nonumber\\
V(\Phi,\eta)&=\lambda^{\Phi \eta}(\Phi^{\dagger}\Phi)\eta^{*}\eta,\quad
V(\Phi,\kappa)=\lambda^{\Phi \kappa}(\Phi^{\dagger}\Phi)\kappa^{*}\kappa,\quad
V(\Phi,\xi)=\lambda^{\Phi \xi}(\Phi^{\dagger}\Phi)\xi^{*}\xi,\nonumber\\
V(\Phi,{\xi^{\prime}})&=\lambda^{\Phi {\xi^{\prime}}}(\Phi^{\dagger}\Phi){\xi^{\prime}}^{*}{\xi^{\prime}},\quad
V(\Delta,\eta)=\lambda^{\eta \Delta }Tr(\Delta^{\dagger}\Delta)\eta^{*}\eta,\quad
V(\Delta,\kappa)=\lambda^{\kappa \Delta}Tr(\Delta^{\dagger}\Delta)\kappa^{*}\kappa,\nonumber\\
V(\Delta,\xi)&=\lambda^{\Delta \xi}Tr(\Delta^{\dagger}\Delta)\xi^{*}\xi,\quad
V(\Delta,{\xi^{\prime}})=\lambda^{\Delta {\xi^{\prime}}}Tr(\Delta^{\dagger}\Delta){\xi^{\prime}}^{*}{\xi^{\prime}},\nonumber\\
V(\eta,\kappa)&=\lambda^{\eta \kappa}(\eta^{*}\eta \kappa^{*}\kappa +\mu^2_{\eta\kappa}(\eta \kappa+h.c)+(\eta^2 \kappa^2+h.c)),\nonumber\\
V(\eta,\xi)&=\lambda^{\eta\xi}\eta^{*}\eta \xi^{*}\xi,\quad V(\eta,{\xi^{\prime}})=\lambda^{\eta{\xi^{\prime}}}\eta^{*}\eta {\xi^{\prime}}^{*}{\xi^{\prime}},\quad
V(\xi,{\xi^{\prime}})=V(\eta, \kappa \rightarrow \xi, {\xi^{\prime}}),\nonumber
\\V(\kappa,\xi)&=\lambda^{\kappa\xi}\kappa^{*}\kappa \xi^{*}\xi,\quad V(\kappa,{\xi^{\prime}})=\lambda^{\kappa{\xi^{\prime}}}\kappa^{*}\kappa {\xi^{\prime}}^{*}{\xi^{\prime}},\quad
V(\phi,\eta,\kappa)=\lambda^{\phi \eta \kappa}((\phi^{\dagger}\phi)\eta \kappa+h.c),\nonumber\\
V(\phi,\xi,{\xi^{\prime}})&=\lambda^{\phi \xi {\xi^{\prime}}}((\phi^{\dagger}\phi)\xi {\xi^{\prime}}+h.c),\quad
V(\Phi,\eta,\kappa)=\lambda^{\Phi \eta \kappa}((\Phi^{\dagger}\Phi)\eta \kappa+h.c),\nonumber\\
V(\Phi,\xi,{\xi^{\prime}})&=\lambda^{\Phi \xi {\xi^{\prime}}}((\Phi^{\dagger}\Phi)\xi {\xi^{\prime}}+h.c),\quad
V(\Delta,\eta,\kappa)=\lambda^{\Delta \eta \kappa}(Tr(\Delta^{\dagger}\Delta)\eta \kappa+h.c),\nonumber\\
V(\Delta,\xi,{\xi^{\prime}})&=\lambda^{\Delta \xi {\xi^{\prime}}}(Tr(\Delta^{\dagger}\Delta)\xi {\xi^{\prime}}+h.c),\quad
V(\eta,\kappa,\xi,{\xi^{\prime}})=\lambda^{\eta \kappa \xi {\xi^{\prime}}}(\eta \kappa \xi {\xi^{\prime}}+h.c).
\end{align}

Many terms containing $(\phi^{\dagger}\phi)_{3_a},\ (\Phi^{\dagger}\Phi)_{3_a}$ and $Tr(\Delta^{\dagger}\Delta)_{3_a}$ are also invariant but are not appeared in \ref{eq:x1} because they are vanished due to the antisymmetric of $3_a$ as a consequence of the tensor product of $3\times3$ of $A_4$ and the VEV alignment of $\phi$, $\Phi$ and $\Delta$.
The minimization conditions (VEVs) of this potential can result in the extreme solutions detailed in
\ref{eq:vacuum}. These VEVs provide information about the stable configurations of the system since they represent crucial places where potential energy is minimized.
\begin{equation}
\label{eq:vacuum}
\begin{aligned}
\left\langle \phi \right\rangle &= (\langle \phi_1\rangle,\quad 0,\quad 0), \quad \langle \phi_1\rangle=(0 \quad v)^{T},\quad
\langle \Phi \rangle = (0,\quad \langle \Phi_2\rangle,\quad \langle \Phi_3 \rangle),\\ \quad \langle \Phi_2\rangle &= \langle \Phi_3\rangle=(u \quad 0)^{T},\quad
\langle \Delta \rangle = (0,\quad \langle \Delta_2\rangle,\quad \langle \Delta_3 \rangle),\\ \quad \langle \Delta_2\rangle &= 
\begin{pmatrix}
    0 & 0 \\
       w & 0\\
  \end{pmatrix}, \quad
\langle \Delta_3\rangle= -\begin{pmatrix}
    0 & 0 \\
       w & 0\\
  \end{pmatrix}
, \quad \left\langle \eta \right\rangle =\left\langle \kappa \right\rangle=v_m, \quad \left\langle \xi \right\rangle =\left\langle \xi^{\prime} \right\rangle=v_\epsilon, 
\end{aligned}
\end{equation}

\textbf{PSO Treatment for Scalar Potential:}  

\begin{figure}
    \centering
    \includegraphics[width=0.9\linewidth]{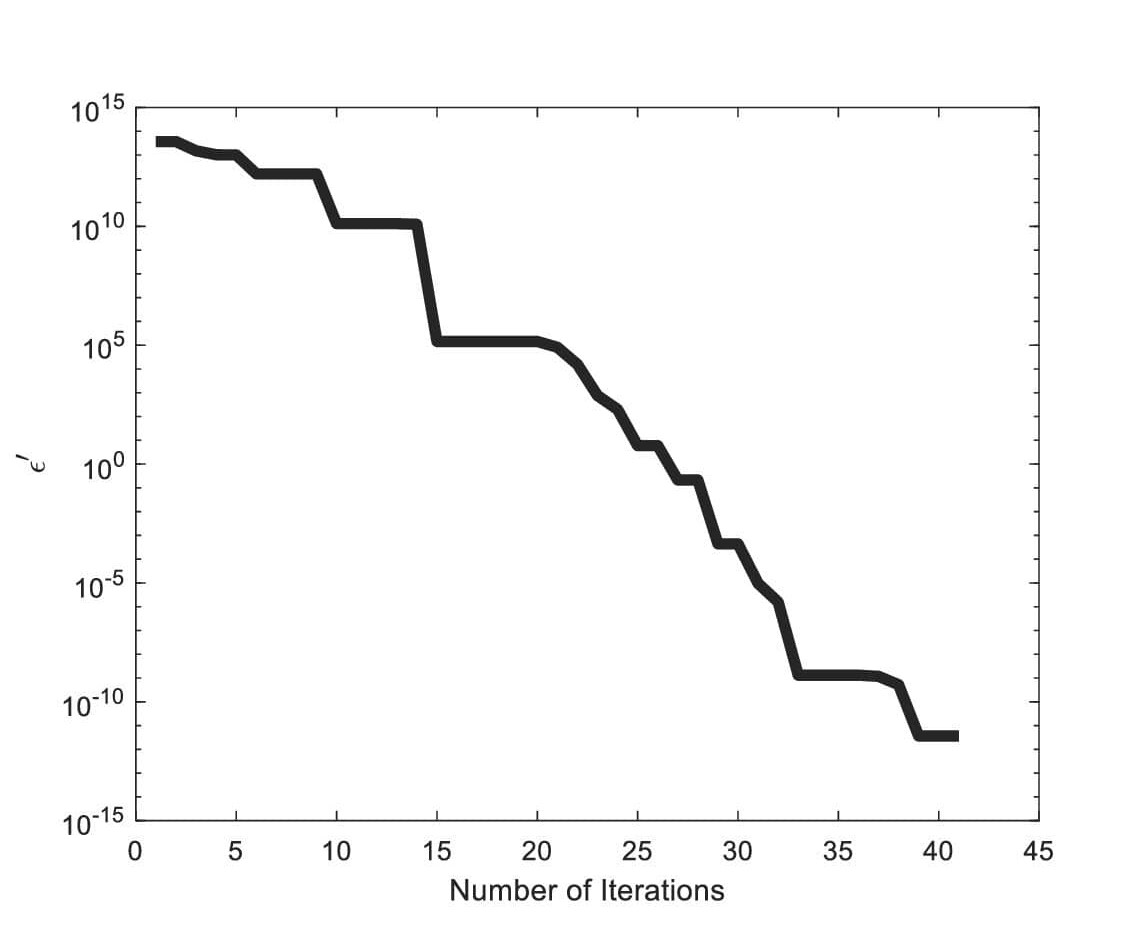}
    \caption{Fitness function of scalar potential versus number of iterations}
    \label{fig:fitunscalarpotential}
\end{figure}

Due to VEVs \ref{eq:vacuum}, multiplication rules for $A_4$ symmetry in the T-diagonal basis \cite{bookfordiscretesymmetries} and $S_2$ symmetry, equation \ref{eq:x1} become as
\begin{equation}
\label{eq:vacuum after sol}
\begin{aligned}
V=&-\mu^2_{\phi}v^2+\lambda^{\phi}_{1}v^4+\frac{4}{9}\lambda^{\phi}_{3}v^4-2\mu^2_{\Phi}u^2+4\lambda^{\Phi}_{1}u^4+\lambda^{\Phi}_{2}u^4+\frac{12}{9}\lambda^{\Phi}_{3}u^4+2\mu^2_{\Delta}w^2+4\lambda^{\Delta}_1w^4\\&+\lambda^{\Delta}_2w^4+\frac{12}{9}\lambda^{\Delta}_3w^4+2\lambda^{\phi\Phi}_{1}v^2u^2-\frac{4}{9}\lambda^{\phi\Phi}_{3}v^2u^2-2\lambda^{\phi\Delta}_1v^2w^2+\frac{4}{9}\lambda^{\phi\Delta}_3v^2w^2-4\lambda^{\Phi\Delta}_1u^2w^2\\&+2\lambda^{\Phi\Delta}_2u^2w^2+\frac{4}{9}\lambda^{\Phi\Delta}_3u^2w^2-\mu^2_\eta v^2_m+\lambda^{\eta}v^4_m-\mu^2_\kappa v^2_m+\lambda^{\kappa}v^4_m-\mu^2_\xi v^2_\epsilon+\lambda^{\xi}v^4_\epsilon-\mu^2_{\xi^{\prime}} v^2_\epsilon\\&+\lambda^{{\xi^{\prime}}}v^4_\epsilon+\lambda^{\phi\eta}v^2v^2_m+\lambda^{\phi\kappa}v^2v^2_m+\lambda^{\phi\xi}v^2v^2_\epsilon+\lambda^{\phi\xi^{\prime}}v^2v^2_\epsilon+2\lambda^{\Phi\eta}u^2v^2_m+2\lambda^{\Phi\kappa}u^2v^2_m\\&+2\lambda^{\Phi\xi}u^2v^2_\epsilon+2\lambda^{\Phi\xi^{\prime}}u^2v^2_\epsilon-2\lambda^{\eta\Delta}w^2v^2_m-2\lambda^{\kappa\Delta}w^2v^2_m-2\lambda^{\Delta\xi}w^2v^2_\epsilon-2\lambda^{\Delta\xi^{\prime}}w^2v^2_\epsilon\\&+\lambda^{\eta\kappa}(v^4_m+2\mu^2_{\eta\kappa}v^2_m+2v^4_m)+\lambda^{\eta\xi}v^2_m v^2_\epsilon+\lambda^{\eta\xi^{\prime}}v^2_m v^2_\epsilon+\lambda^{\xi\xi^{\prime}}(v^4_\epsilon+2\mu^2_{\xi\xi^{\prime}}v^2_\epsilon+2v^4_\epsilon)\\&+\lambda^{\kappa\xi}v^2_mv^2_\epsilon+\lambda^{\kappa\xi^{\prime}}v^2_m v^2_\epsilon+2\lambda^{\phi\eta\kappa}v^2 v^2_m+2\lambda^{\phi\xi\xi^{\prime}}v^2 v^2_\epsilon+4\lambda^{\Phi\eta\kappa}u^2v^2_m+4\lambda^{\Phi\xi\xi^{\prime}}u^2v^2_\epsilon\\&-4\lambda^{\Delta\eta\kappa}w^2v^2_m-4\lambda^{\Delta\xi\xi^{\prime}}w^2v^2_\epsilon+2\lambda^{\eta\kappa\xi\xi^{\prime}}v^2_mv^2_\epsilon,
\end{aligned}
\end{equation}
and the fitness function $(\epsilon^{\prime})$ for \ref{eq:vacuum after sol} is expressed as follows.
\begin{equation}
\label{eq:objectivefunctionforpotential}
\begin{aligned}
\epsilon^{\prime}=\biggl[&-\mu^2_{\phi}v^2+\lambda^{\phi}_{1}v^4+\frac{4}{9}\lambda^{\phi}_{3}v^4-2\mu^2_{\Phi}u^2+4\lambda^{\Phi}_{1}u^4+\lambda^{\Phi}_{2}u^4+\frac{12}{9}\lambda^{\Phi}_{3}u^4+2\mu^2_{\Delta}w^2+4\lambda^{\Delta}_1w^4\\&+\lambda^{\Delta}_2w^4+\frac{12}{9}\lambda^{\Delta}_3w^4+2\lambda^{\phi\Phi}_{1}v^2u^2-\frac{4}{9}\lambda^{\phi\Phi}_{3}v^2u^2-2\lambda^{\phi\Delta}_1v^2w^2+\frac{4}{9}\lambda^{\phi\Delta}_3v^2w^2-4\lambda^{\Phi\Delta}_1u^2w^2\\&+2\lambda^{\Phi\Delta}_2u^2w^2+\frac{4}{9}\lambda^{\Phi\Delta}_3u^2w^2-\mu^2_\eta v^2_m+\lambda^{\eta}v^4_m-\mu^2_\kappa v^2_m+\lambda^{\kappa}v^4_m-\mu^2_\xi v^2_\epsilon+\lambda^{\xi}v^4_\epsilon-\mu^2_{\xi^{\prime}} v^2_\epsilon\\&+\lambda^{{\xi^{\prime}}}v^4_\epsilon+\lambda^{\phi\eta}v^2v^2_m+\lambda^{\phi\kappa}v^2v^2_m+\lambda^{\phi\xi}v^2v^2_\epsilon+\lambda^{\phi\xi^{\prime}}v^2v^2_\epsilon+2\lambda^{\Phi\eta}u^2v^2_m+2\lambda^{\Phi\kappa}u^2v^2_m\\&+2\lambda^{\Phi\xi}u^2v^2_\epsilon+2\lambda^{\Phi\xi^{\prime}}u^2v^2_\epsilon-2\lambda^{\eta\Delta}w^2v^2_m-2\lambda^{\kappa\Delta}w^2v^2_m-2\lambda^{\Delta\xi}w^2v^2_\epsilon-2\lambda^{\Delta\xi^{\prime}}w^2v^2_\epsilon\\&+\lambda^{\eta\kappa}(v^4_m+2\mu^2_{\eta\kappa}v^2_m+2v^4_m)+\lambda^{\eta\xi}v^2_m v^2_\epsilon+\lambda^{\eta\xi^{\prime}}v^2_m v^2_\epsilon+\lambda^{\xi\xi^{\prime}}(v^4_\epsilon+2\mu^2_{\xi\xi^{\prime}}v^2_\epsilon+2v^4_\epsilon)\\&+\lambda^{\kappa\xi}v^2_mv^2_\epsilon+\lambda^{\kappa\xi^{\prime}}v^2_m v^2_\epsilon+2\lambda^{\phi\eta\kappa}v^2 v^2_m+2\lambda^{\phi\xi\xi^{\prime}}v^2 v^2_\epsilon+4\lambda^{\Phi\eta\kappa}u^2v^2_m+4\lambda^{\Phi\xi\xi^{\prime}}u^2v^2_\epsilon\\&-4\lambda^{\Delta\eta\kappa}w^2v^2_m-4\lambda^{\Delta\xi\xi^{\prime}}w^2v^2_\epsilon+2\lambda^{\eta\kappa\xi\xi^{\prime}}v^2_mv^2_\epsilon\biggl]^{2},
\end{aligned}
\end{equation}
To stimulate the advancement of meta-heuristic optimization algorithms, we again utilized PSO technique to minimize the scalar potential. The objective function is minimized through PSO for scalar potential with 500 iteration is presented in figure \ref{fig:fitunscalarpotential}.
This figure demonstrates that the objective function converges to zero with each iteration when employing the VEVs provided in \ref{eq:vacuum}. The optimal parameter values, as measured within the objective function of the scalar potential using PSO technique, are listed in table
\ref{tab:optimal values for potential analysis}.
\begin{table}[htbp]
\centering
\begin{tabular}{llll}
\hline
 Parameters& Optimal values & Parameters &Optimal values
 \\
\hline
$\mu_\phi$ &$179.175 \ GeV$  & $\mu_\Phi$& $139.331 \ GeV$ \\
\hline
$\mu_\Delta$ &$145.538 \ GeV$ &  $\mu_\eta$& $178.33 \ GeV$\\
\hline
$\mu_\kappa$ & $115.583 \ GeV$ & $\mu_\xi$&$187.052 \ GeV$  \\
\hline
$\mu_{\xi^{\prime}}$ &$153.417 \ GeV$  & $\mu_{\eta\kappa}$&$104.519 \ GeV$  \\
\hline
$\mu_{\xi \xi^{\prime}}$ & $158.023 \ GeV$ & $v$& $134.316 \ GeV$ \\
\hline
$u$ & $113.798 \ GeV$ & $w$& $134.684 \ GeV$ \\
\hline
$v_m$ & $158.347 \ GeV$ & $v_\epsilon$&  $159.224 \ GeV$\\
\hline
$\lambda^{\phi}_1$ & $10.00000$ &$\lambda^{\phi}_3$ &  $6.08711$
\\
\hline
$\lambda^{\Phi}_1$ & $-8.95786$ &$\lambda^{\Phi}_2$ &  $3.55938$\\
\hline
 $\lambda^{\Phi}_3$& $-2.54363$ &$\lambda^{\Delta}_1$& $-2.6438$ \\
\hline
 $\lambda^{\Delta}_2$ &  $3.96077$&$\lambda^{\Delta}_3$ & $2.64742$ \\
\hline
$\lambda^{\phi\Phi}_1$ &  $9.7041$& $\lambda^{\phi\Phi}_3$&  $0.225504$\\
\hline
$\lambda^{\phi\Delta}_1$ & $7.95572$ &$\lambda^{\phi\Delta}_3$& $-6.47177$ \\
\hline
$\lambda^{\Phi\Delta}_1$ & $-0.953689$ &$\lambda^{\Phi\Delta}_2 $&  $3.49581$\\
\hline
$\lambda^{\Phi\Delta}_3$ & $-5.0444$ &$\lambda^\eta $ &  $-10.00000$\\
\hline
$\lambda^\kappa $  & $9.14604$ &$\lambda^\xi$& $-2.93147$ \\
\hline
$\lambda^{\xi^{\prime}}$ & $-2.35666$ &$\lambda^{\phi\eta}$ & $10.00000$ \\
\hline
 $\lambda^{\phi\kappa}$& $8.98456$ &$\lambda^{\phi\xi}$ &  $-10.00000$\\
\hline
$\lambda^{\phi\xi^{\prime}}$ & $1.73548$ &$\lambda^{\Phi\eta}$ & $-5.4436$ \\
\hline
$\lambda^{\Phi\kappa}$ & $-4.29061$ &$\lambda^{\Phi\xi}$ & $-3.53999$ \\
\hline
 $\lambda^{\Phi\xi^{\prime}}$& $1.25844$ & $\lambda^{\eta\Delta}$& $-4.9442$ \\
\hline
$\lambda^{\kappa\Delta}$ & $-9.23625$ &$\lambda^{\Delta\xi}$& $-2.61907$  \\
\hline
$\lambda^{\Delta\xi^{\prime}}$ & $-5.5742$ &$\lambda^{\eta\kappa}$& $10.00000$ \\
\hline
 $\lambda^{\eta\xi}$& $-1.84958$ & $\lambda^{\eta\xi^{\prime}}$&  $2.21404$\\
\hline
 $\lambda^{\xi\xi^{\prime}}$& $2.37551$ &$\lambda^{\kappa\xi}$ & $10.00000$ \\
\hline
 $\lambda^{\kappa\xi^{\prime}}$& $-7.90574$ &$\lambda^{\phi\eta\kappa}$ &  $-10.00000$\\
\hline
$\lambda^{\phi\xi\xi^{\prime}}$ & $-0.973407$ &$\lambda^{\Phi\eta\kappa}$ & $-2.65544$ \\
\hline
$\lambda^{\Phi\xi\xi^{\prime}}$ & $-10.00000$ & $\lambda^{\Delta\eta\kappa}$&  $-1.32635$\\
\hline
$\lambda^{\Delta\xi\xi^{\prime}}$ & $7.21468$ & $\lambda^{\eta\kappa\xi\xi^{\prime}}$& 1.23132 \\
\hline
\end{tabular}
\caption{The optimal values of parameters given in \ref{eq:x1} through PSO.
\label{tab:optimal values for potential analysis}}
\end{table}
The scalar potential is minimized from these optimal values. In contrast to \cite{Model2023}, which analyzed the scalar potential without determining the masses of extra flavon fields, the PSO approach not only analyzes the scalar potential but also determines the numerical values of the flavon field masses. Table \ref{tab:optimal values for potential analysis}, revealing that the flavon masses are on the order of $O(100\ GeV)$, which is relevant to current experimental searches.
\section{Conclusion}
\label{sec:Conclusion}
In this study, we have examined a model within $SU(2)_L \times U(1)_Y \times A_{4}\times S_2\times Z_{10} \times Z_{3}$ to estimate the neutrino masses using particle swarm optimization technique for both neutrino hierarchy. In this model, a hybrid seesaw mechanism proposed for improved mass suppression and new mixing patterns by combining type-I and type-II and generate effective Majorana neutrino mass matrices. After calculating the mass eigenvalues and lepton mixing matrix upto second order perturbation theory in the framework based on $A_{4}$ symmetry, we investigated the minimization of the scalar potential for VEVs through PSO. The utilization of PSO in determining optimal parameters for computing $U_{PMNS}$ matrices, neutrino masses as: ${m_{1}^{\prime}}^{(N)}(upper)=4.0000 \times 10^{-2}\ eV$, ${m_{2}^{\prime}}^{(N)}(upper)=4.0000 \times 10^{-2}\ eV$,  
${m_{3}^{\prime}}^{(N)}(upper)=4.0000 \times 10^{-2}\ eV$, ${m_{1}^{\prime}}^{(I)}(upper)=3.8628\times 10^{-2}\ eV$,
${m_{2}^{\prime}}^{(I)}(upper)=4.0548\times 10^{-2}\ eV$,
${m_{3}^{\prime}}^{(I)}(upper)=3.8532\times 10^{-2}\ eV$,
${m_{1}^{\prime}}^{(N)}(lower)=2.0000 \times 10^{-2}\ eV$, 
${m_{2}^{\prime}}^{(N)}(lower)=2.0000 \times 10^{-2}\ eV$, 
${m_{3}^{\prime}}^{(N)}(lower)=2.0000 \times 10^{-2}\ eV$, 
${m_{1}^{\prime}}^{(I)}(lower)=1.1049\times 10^{-2}\ eV$,
${m_{2}^{\prime}}^{(I)}(lower)=3.9298\times 10^{-2}\ eV$ and
${m_{3}^{\prime}}^{(I)}(lower)=9.6381\times 10^{-3}\ eV$, effective neutrino mass parameters as: 
$\langle {m_{ee}} \rangle^{N}(upper)=40.0050 \ meV$, $\langle {m_{\beta}} \rangle^{N}(upper)=40.0025\ meV$, $\langle {m_{ee}} \rangle^{I}(upper)=39.2181\ meV$, $\langle {m_{\beta}} \rangle^{I}(upper)=39.2257\ meV$, $\langle {m_{ee}} \rangle^{N}(lower)=20.0024\ meV$, $\langle {m_{\beta}} \rangle^{N}(lower)=20.0012\ meV$, $\langle {m_{ee}} \rangle^{I}(lower)=19.6608\ meV$, $\langle {m_{\beta}} \rangle^{I}(lower)=23.5908\ meV$. for both mass hierarchy are illustrated as well. The Dirac CP-violating phase $\delta$ is measured to be $-\pi/2$.
\section*{Data availability}
The datasets used and/or analysed during the current study available from the corresponding author on reasonable request.

\bibliographystyle{JHEP}
\bibliography{biblio}

@article{kamiokande1998evidence,
author = {{Fukuda}, Y. and {Hayakawa}, T. and {Ichihara}, E. and {Inoue}, K. and {Ishihara}, K. and {Ishino}, H. and {Itow}, Y. and {Kajita}, T. and {Kameda}, J. and {Kasuga}, S. and {Kobayashi}, K. and {Kobayashi}, Y. and {Koshio}, Y. and {Miura}, M. and {Nakahata}, M. and {Nakayama}, S. and {Okada}, A. and {Okumura}, K. and {Sakurai}, N. and {Shiozawa}, M. and {Suzuki}, Y. and {Takeuchi}, Y. and {Totsuka}, Y. and {Yamada}, S. and {Earl}, M. and {Habig}, A. and {Kearns}, E. and {Messier}, M.~D. and {Scholberg}, K. and {Stone}, J.~L. and {Sulak}, L.~R. and {Walter}, C.~W. and {Goldhaber}, M. and {Barszczxak}, T. and {Casper}, D. and {Gajewski}, W. and {Halverson}, P.~G. and {Hsu}, J. and {Kropp}, W.~R. and {Price}, L.~R. and {Reines}, F. and {Smy}, M. and {Sobel}, H.~W. and {Vagins}, M.~R. and {Ganezer}, K.~S. and {Keig}, W.~E. and {Ellsworth}, R.~W. and {Tasaka}, S. and {Flanagan}, J.~W. and {Kibayashi}, A. and {Learned}, J.~G. and {Matsuno}, S. and {Stenger}, V.~J. and {Takemori}, D. and {Ishii}, T. and {Kanzaki}, J. and {Kobayashi}, T. and {Mine}, S. and {Nakamura}, K. and {Nishikawa}, K. and {Oyama}, Y. and {Sakai}, A. and {Sakuda}, M. and {Sasaki}, O. and {Echigo}, S. and {Kohama}, M. and {Suzuki}, A.~T. and {Haines}, T.~J. and {Blaufuss}, E. and {Kim}, B.~K. and {Sanford}, R. and {Svoboda}, R. and {Chen}, M.~L. and {Conner}, Z. and {Goodman}, J.~A. and {Sullivan}, G.~W. and {Hill}, J. and {Jung}, C.~K. and {Martens}, K. and {Mauger}, C. and {McGrew}, C. and {Sharkey}, E. and {Viren}, B. and {Yanagisawa}, C. and {Doki}, W. and {Miyano}, K. and {Okazawa}, H. and {Saji}, C. and {Takahata}, M. and {Nagashima}, Y. and {Takita}, M. and {Yamaguchi}, T. and {Yoshida}, M. and {Kim}, S.~B. and {Etoh}, M. and {Fujita}, K. and {Hasegawa}, A. and {Hasegawa}, T. and {Hatakeyama}, S. and {Iwamoto}, T. and {Koga}, M. and {Maruyama}, T. and {Ogawa}, H. and {Shirai}, J. and {Suzuki}, A. and {Tsushima}, F. and {Koshiba}, M. and {Nemoto}, M. and {Nishijima}, K. and {Futagami}, T. and {Hayato}, Y. and {Kanaya}, Y. and {Kaneyuki}, K. and {Watanabe}, Y. and {Kielczewska}, D. and {Doyle}, R.~A. and {George}, J.~S. and {Stachyra}, A.~L. and {Wai}, L.~L. and {Wilkes}, R.~J. and {Young}, K.~K.},
        title = "{Evidence for Oscillation of Atmospheric Neutrinos}",
      journal = {Physical Review Letters},
         year = 1998,
        month = aug,
       volume = {81},
       number = {8},
        pages = {1562-1567},
          doi = {10.1103/PhysRevLett.81.1562},
archivePrefix = {arXiv},
       eprint = {hep-ex/9807003},
 primaryClass = {hep-ex},
       adsurl = {https://ui.adsabs.harvard.edu/abs/1998PhRvL..81.1562F}
}

@article{eguchi2003first,
author = {{Eguchi}, K. and {Enomoto}, S. and {Furuno}, K. and {Goldman}, J. and {Hanada}, H. and {Ikeda}, H. and {Ikeda}, K. and {Inoue}, K. and {Ishihara}, K. and {Itoh}, W. and {Iwamoto}, T. and {Kawaguchi}, T. and {Kawashima}, T. and {Kinoshita}, H. and {Kishimoto}, Y. and {Koga}, M. and {Koseki}, Y. and {Maeda}, T. and {Mitsui}, T. and {Motoki}, M. and {Nakajima}, K. and {Nakajima}, M. and {Nakajima}, T. and {Ogawa}, H. and {Owada}, K. and {Sakabe}, T. and {Shimizu}, I. and {Shirai}, J. and {Suekane}, F. and {Suzuki}, A. and {Tada}, K. and {Tajima}, O. and {Takayama}, T. and {Tamae}, K. and {Watanabe}, H. and {Busenitz}, J. and {Djurcic}, Z. and {McKinny}, K. and {Mei}, D. -M. and {Piepke}, A. and {Yakushev}, E. and {Berger}, B.~E. and {Chan}, Y.~D. and {Decowski}, M.~P. and {Dwyer}, D.~A. and {Freedman}, S.~J. and {Fu}, Y. and {Fujikawa}, B.~K. and {Heeger}, K.~M. and {Lesko}, K.~T. and {Luk}, K. -B. and {Murayama}, H. and {Nygren}, D.~R. and {Okada}, C.~E. and {Poon}, A.~W. and {Steiner}, H.~M. and {Winslow}, L.~A. and {Horton-Smith}, G.~A. and {McKeown}, R.~D. and {Ritter}, J. and {Tipton}, B. and {Vogel}, P. and {Lane}, C.~E. and {Miletic}, T. and {Gorham}, P.~W. and {Guillian}, G. and {Learned}, J.~G. and {Maricic}, J. and {Matsuno}, S. and {Pakvasa}, S. and {Dazeley}, S. and {Hatakeyama}, S. and {Murakami}, M. and {Svoboda}, R.~C. and {Dieterle}, B.~D. and {Dimauro}, M. and {Detwiler}, J. and {Gratta}, G. and {Ishii}, K. and {Tolich}, N. and {Uchida}, Y. and {Batygov}, M. and {Bugg}, W. and {Cohn}, H. and {Efremenko}, Y. and {Kamyshkov}, Y. and {Kozlov}, A. and {Nakamura}, Y. and {de Braeckeleer}, L. and {Gould}, C.~R. and {Karwowski}, H.~J. and {Markoff}, D.~M. and {Messimore}, J.~A. and {Nakamura}, K. and {Rohm}, R.~M. and {Tornow}, W. and {Young}, A.~R. and {Wang}, Y. -F.},
        title = "{First Results from KamLAND: Evidence for Reactor Antineutrino Disappearance}",
      journal = {Physical Review Letters},
         year = 2003,
        month = jan,
       volume = {90},
       number = {2},
          eid = {021802},
        pages = {021802},
          doi = {10.1103/PhysRevLett.90.021802},
archivePrefix = {arXiv},
       eprint = {hep-ex/0212021},
 primaryClass = {hep-ex},
       adsurl = {https://ui.adsabs.harvard.edu/abs/2003PhRvL..90b1802E}
}

@article{ahn2006measurement,
  author = {{Ahn}, M.~H. and {Aliu}, E. and {Andringa}, S. and {Aoki}, S. and {Aoyama}, Y. and {Argyriades}, J. and {Asakura}, K. and {Ashie}, R. and {Berghaus}, F. and {Berns}, H.~G. and {Bhang}, H. and {Blondel}, A. and {Borghi}, S. and {Bouchez}, J. and {Boyd}, S.~C. and {Burguet-Castell}, J. and {Casper}, D. and {Catala}, J. and {Cavata}, C. and {Cervera}, A. and {Chen}, S.~M. and {Cho}, K.~O. and {Choi}, J.~H. and {Dore}, U. and {Echigo}, S. and {Espinal}, X. and {Fechner}, M. and {Fernandez}, E. and {Fujii}, K. and {Fujii}, Y. and {Fukuda}, S. and {Fukuda}, Y. and {Gomez-Cadenas}, J. and {Gran}, R. and {Hara}, T. and {Hasegawa}, M. and {Hasegawa}, T. and {Hayashi}, K. and {Hayato}, Y. and {Helmer}, R.~L. and {Higuchi}, I. and {Hill}, J. and {Hiraide}, K. and {Hirose}, E. and {Hosaka}, J. and {Ichikawa}, A.~K. and {Ieiri}, M. and {Iinuma}, M. and {Ikeda}, A. and {Inagaki}, T. and {Ishida}, T. and {Ishihara}, K. and {Ishii}, H. and {Ishii}, T. and {Ishino}, H. and {Ishitsuka}, M. and {Itow}, Y. and {Iwashita}, T. and {Jang}, H.~I. and {Jang}, J.~S. and {Jeon}, E.~J. and {Jeong}, I.~S. and {Joo}, K.~K. and {Jover}, G. and {Jung}, C.~K. and {Kajita}, T. and {Kameda}, J. and {Kaneyuki}, K. and {Kang}, B.~H. and {Kato}, I. and {Kato}, Y. and {Kearns}, E. and {Kerr}, D. and {Kim}, C.~O. and {Khabibullin}, M. and {Khotjantsev}, A. and {Kielczewska}, D. and {Kim}, B.~J. and {Kim}, H.~I. and {Kim}, J.~H. and {Kim}, J.~Y. and {Kim}, S.~B. and {Kitamura}, M. and {Kitching}, P. and {Kobayashi}, K. and {Kobayashi}, T. and {Kohama}, M. and {Konaka}, A. and {Koshio}, Y. and {Kropp}, W. and {Kubota}, J. and {Kudenko}, Yu. and {Kume}, G. and {Kuno}, Y. and {Kurimoto}, Y. and {Kutter}, T. and {Learned}, J. and {Likhoded}, S. and {Lim}, I.~T. and {Lim}, S.~H. and {Loverre}, P.~F. and {Ludovici}, L. and {Maesaka}, H. and {Mallet}, J. and {Mariani}, C. and {Martens}, K. and {Maruyama}, T. and {Matsuno}, S. and {Matveev}, V. and {Mauger}, C. and {McConnel Mahn}, K.~B. and {McGrew}, C. and {Mikheyev}, S. and {Minakawa}, M. and {Minamino}, A. and {Mine}, S. and {Mineev}, O. and {Mitsuda}, C. and {Mitsuka}, G. and {Miura}, M. and {Moriguchi}, Y. and {Morita}, T. and {Moriyama}, S. and {Nakadaira}, T. and {Nakahata}, M. and {Nakamura}, K. and {Nakano}, I. and {Nakata}, F. and {Nakaya}, T. and {Nakayama}, S. and {Namba}, T. and {Nambu}, R. and {Nawang}, S. and {Nishikawa}, K. and {Nishino}, H. and {Nishiyama}, S. and {Nitta}, K. and {Noda}, S. and {Noumi}, H. and {Nova}, F. and {Novella}, P. and {Obayashi}, Y. and {Okada}, A. and {Okumura}, K. and {Okumura}, M. and {Onchi}, M. and {Ooyabu}, T. and {Oser}, S.~M. and {Otaki}, T. and {Oyama}, Y. and {Pac}, M.~Y. and {Park}, H. and {Pierre}, F. and {Rodriguez}, A. and {Saji}, C. and {Sakai}, A. and {Sakuda}, M. and {Sakurai}, N. and {Sanchez}, F. and {Sarrat}, A. and {Sasaki}, T. and {Sato}, H. and {Sato}, K. and {Scholberg}, K. and {Schroeter}, R. and {Sekiguchi}, M. and {Seo}, E. and {Sharkey}, E. and {Shima}, A. and {Shiozawa}, M. and {Shiraishi}, K. and {Sitjes}, G. and {Smy}, M. and {So}, H. and {Sobel}, H. and {Sorel}, M. and {Stone}, J. and {Sulak}, L. and {Suga}, Y. and {Suzuki}, A. and {Suzuki}, Y. and {Suzuki}, Y. and {Tada}, M. and {Takahashi}, T. and {Takasaki}, M. and {Takatsuki}, M. and {Takenaga}, Y. and {Takenaka}, K. and {Takeuchi}, H. and {Takeuchi}, Y. and {Taki}, K. and {Takubo}, Y. and {Tamura}, N. and {Tanaka}, H. and {Tanaka}, K. and {Tanaka}, M. and {Tanaka}, Y. and {Tashiro}, K. and {Terri}, R. and {T'Jampens}, S. and {Tornero-Lopez}, A. and {Toshito}, T. and {Totsuka}, Y. and {Ueda}, S. and {Vagins}, M. and {Whitehead}, L. and {Walter}, C.~W. and {Wang}, W. and {Wilkes}, R.~J. and {Yamada}, S. and {Yamada}, Y. and {Yamamoto}, S. and {Yamanoi}, Y. and {Yanagisawa}, C. and {Yershov}, N. and {Yokoyama}, H. and {Yokoyama}, M. and {Yoo}, J. and {Yoshida}, M. and {Zalipska}, J.},
        title = "{Measurement of neutrino oscillation by the K2K experiment}",
      journal = {Physical Review D},
         year = 2006,
        month = oct,
       volume = {74},
       number = {7},
          eid = {072003},
        pages = {072003},
          doi = {10.1103/PhysRevD.74.072003},
archivePrefix = {arXiv},
       eprint = {hep-ex/0606032},
 primaryClass = {hep-ex},
       adsurl = {https://ui.adsabs.harvard.edu/abs/2006PhRvD..74g2003A}
}

@article{michael2006observation,
  author = {{Michael}, D.~G. and {Adamson}, P. and {Alexopoulos}, T. and {Allison}, W.~W.~M. and {Alner}, G.~J. and {Anderson}, K. and {Andreopoulos}, C. and {Andrews}, M. and {Andrews}, R. and {Arms}, K.~E. and {Armstrong}, R. and {Arroyo}, C. and {Auty}, D.~J. and {Avvakumov}, S. and {Ayres}, D.~S. and {Baller}, B. and {Barish}, B. and {Barker}, M.~A. and {Barnes}, P.~D., Jr. and {Barr}, G. and {Barrett}, W.~L. and {Beall}, E. and {Becker}, B.~R. and {Belias}, A. and {Bergfeld}, T. and {Bernstein}, R.~H. and {Bhattacharya}, D. and {Bishai}, M. and {Blake}, A. and {Bocean}, V. and {Bock}, B. and {Bock}, G.~J. and {Boehm}, J. and {Boehnlein}, D.~J. and {Bogert}, D. and {Border}, P.~M. and {Bower}, C. and {Boyd}, S. and {Buckley-Geer}, E. and {Bungau}, C. and {Byon-Wagner}, A. and {Cabrera}, A. and {Chapman}, J.~D. and {Chase}, T.~R. and {Cherdack}, D. and {Chernichenko}, S.~K. and {Childress}, S. and {Choudhary}, B.~C. and {Cobb}, J.~H. and {Cossairt}, J.~D. and {Courant}, H. and {Crane}, D.~A. and {Culling}, A.~J. and {Dawson}, J.~W. and {de Jong}, J.~K. and {Demuth}, D.~M. and {de Santo}, A. and {Dierckxsens}, M. and {Diwan}, M.~V. and {Dorman}, M. and {Drake}, G. and {Drakoulakos}, D. and {Ducar}, R. and {Durkin}, T. and {Erwin}, A.~R. and {Escobar}, C.~O. and {Evans}, J.~J. and {Fackler}, O.~D. and {Falk Harris}, E. and {Feldman}, G.~J. and {Felt}, N. and {Fields}, T.~H. and {Ford}, R. and {Frohne}, M.~V. and {Gallagher}, H.~R. and {Gebhard}, M. and {Giurgiu}, G.~A. and {Godley}, A. and {Gogos}, J. and {Goodman}, M.~C. and {Gornushkin}, Yu. and {Gouffon}, P. and {Gran}, R. and {Grashorn}, E. and {Grossman}, N. and {Grudzinski}, J.~J. and {Grzelak}, K. and {Guarino}, V. and {Habig}, A. and {Halsall}, R. and {Hanson}, J. and {Harris}, D. and {Harris}, P.~G. and {Hartnell}, J. and {Hartouni}, E.~P. and {Hatcher}, R. and {Heller}, K. and {Hill}, N. and {Ho}, Y. and {Holin}, A. and {Howcroft}, C. and {Hylen}, J. and {Ignatenko}, M. and {Indurthy}, D. and {Irwin}, G.~M. and {Ishitsuka}, M. and {Jaffe}, D.~E. and {James}, C. and {Jenner}, L. and {Jensen}, D. and {Joffe-Minor}, T. and {Kafka}, T. and {Kang}, H.~J. and {Kasahara}, S.~M.~S. and {Kilmer}, J. and {Kim}, H. and {Kim}, M.~S. and {Koizumi}, G. and {Kopp}, S. and {Kordosky}, M. and {Koskinen}, D.~J. and {Kostin}, M. and {Kotelnikov}, S.~K. and {Krakauer}, D.~A. and {Kreymer}, A. and {Kumaratunga}, S. and {Ladran}, A.~S. and {Lang}, K. and {Laughton}, C. and {Lebedev}, A. and {Lee}, R. and {Lee}, W.~Y. and {Libkind}, M.~A. and {Ling}, J. and {Liu}, J. and {Litchfield}, P.~J. and {Litchfield}, R.~P. and {Longley}, N.~P. and {Lucas}, P. and {Luebke}, W. and {Madani}, S. and {Maher}, E. and {Makeev}, V. and {Mann}, W.~A. and {Marchionni}, A. and {Marino}, A.~D. and {Marshak}, M.~L. and {Marshall}, J.~S. and {Mayer}, N. and {McDonald}, J. and {McGowan}, A.~M. and {Meier}, J.~R. and {Merzon}, G.~I. and {Messier}, M.~D. and {Milburn}, R.~H. and {Miller}, J.~L. and {Miller}, W.~H. and {Mishra}, S.~R. and {Mislivec}, A. and {Miyagawa}, P.~S. and {Moore}, C.~D. and {Morf{\'\i}n}, J. and {Morse}, R. and {Mualem}, L. and {Mufson}, S. and {Murgia}, S. and {Murtagh}, M.~J. and {Musser}, J. and {Naples}, D. and {Nelson}, C. and {Nelson}, J.~K. and {Newman}, H.~B. and {Nezrick}, F. and {Nichol}, R.~J. and {Nicholls}, T.~C. and {Ochoa-Ricoux}, J.~P. and {Oliver}, J. and {Oliver}, W.~P. and {Onuchin}, V.~A. and {Osiecki}, T. and {Ospanov}, R. and {Paley}, J. and {Paolone}, V. and {Para}, A. and {Patzak}, T. and {Pavlovi{\'c}}, {\v{Z}}. and {Pearce}, G.~F. and {Pearson}, N. and {Peck}, C.~W. and {Perry}, C. and {Peterson}, E.~A. and {Petyt}, D.~A. and {Ping}, H. and {Piteira}, R. and {Pittam}, R. and {Pla-Dalmau}, A. and {Plunkett}, R.~K. and {Price}, L.~E. and {Proga}, M. and {Pushka}, D.~R. and {Rahman}, D. and {Rameika}, R.~A. and {Raufer}, T.~M. and {Read}, A.~L. and {Rebel}, B. and {Reichenbacher}, J. and {Reyna}, D.~E. and {Rosenfeld}, C. and {Rubin}, H.~A. and {Ruddick}, K. and {Ryabov}, V.~A. and {Saakyan}, R. and {Sanchez}, M.~C. and {Saoulidou}, N. and {Schneps}, J. and {Schoessow}, P.~V. and {Schreiner}, P. and {Schwienhorst}, R. and {Semenov}, V.~K. and {Seun}, S. -M. and {Shanahan}, P. and {Shield}, P.~D. and {Smart}, W. and {Smirnitsky}, V. and {Smith}, C. and {Smith}, P.~N. and {Sousa}, A. and {Speakman}, B. and {Stamoulis}, P. and {Stefanik}, A. and {Sullivan}, P. and {Swan}, J.~M. and {Symes}, P.~A. and {Tagg}, N. and {Talaga}, R.~L. and {Terekhov}, A. and {Tetteh-Lartey}, E. and {Thomas}, J. and {Thompson}, J. and {Thomson}, M.~A. and {Thron}, J.~L. and {Tinti}, G. and {Trendler}, R. and {Trevor}, J. and {Trostin}, I. and {Tsarev}, V.~A. and {Tzanakos}, G. and {Urheim}, J. and {Vahle}, P. and {Vakili}, M. and {Vaziri}, K. and {Velissaris}, C. and {Verebryusov}, V. and {Viren}, B. and {Wai}, L. and {Ward}, C.~P. and {Ward}, D.~R. and {Watabe}, M. and {Weber}, A. and {Webb}, R.~C. and {Wehmann}, A. and {West}, N. and {White}, C. and {White}, R.~F. and {Wojcicki}, S.~G. and {Wright}, D.~M. and {Wu}, Q.~K. and {Yan}, W.~G. and {Yang}, T. and {Yumiceva}, F.~X. and {Yun}, J.~C. and {Zheng}, H. and {Zois}, M. and {Zwaska}, R.},
        title = "{Observation of Muon Neutrino Disappearance with the MINOS Detectors in the NuMI Neutrino Beam}",
      journal = {Physical Review Letters},
         year = 2006,
        month = nov,
       volume = {97},
       number = {19},
          eid = {191801},
        pages = {191801},
          doi = {10.1103/PhysRevLett.97.191801},
archivePrefix = {arXiv},
       eprint = {hep-ex/0607088},
 primaryClass = {hep-ex},
       adsurl = {https://ui.adsabs.harvard.edu/abs/2006PhRvL..97s1801M}
}

@article{sh2009opera,
author = {{Acquafredda}, R. and {Adam}, T. and {Agafonova}, N. and {Alvarez Sanchez}, P. and {Ambrosio}, M. and {Anokhina}, A. and {Aoki}, S. and {Ariga}, A. and {Ariga}, T. and {Arrabito}, L. and {Aufranc}, C. and {Autiero}, D. and {Badertscher}, A. and {Bagulya}, A. and {Baussan}, E. and {Bergnoli}, A. and {Bersani Greggio}, F. and {Bertolin}, A. and {Besnier}, M. and {Biar{\'e}}, D. and {Bick}, D. and {Blin}, S. and {Borer}, K. and {Boucrot}, J. and {Boutigny}, D. and {Boyarkin}, V. and {Bozza}, C. and {Brugi{\`e}re}, T. and {Brugnera}, R. and {Brunetti}, G. and {Buontempo}, S. and {Campagne}, J.~E. and {Carlus}, B. and {Carrara}, E. and {Cazes}, A. and {Chaussard}, L. and {Chernyavsky}, M. and {Chiarella}, V. and {Chon-Sen}, N. and {Chukanov}, A. and {Ciesielski}, R. and {Consiglio}, L. and {Cozzi}, M. and {D'Amato}, G. and {Dal Corso}, F. and {D'Ambrosio}, N. and {Damet}, J. and {de La Taille}, C. and {DeLellis}, G. and {D{\'e}clais}, Y. and {Descombes}, T. and {DeSerio}, M. and {Di Capua}, F. and {Di Ferdinando}, D. and {Di Giovanni}, A. and {Di Marco}, N. and {Di Troia}, C. and {Dick}, N. and {Dmitrievski}, S. and {Dominjon}, A. and {Dracos}, M. and {Duchesneau}, D. and {Dulach}, B. and {Dusini}, S. and {Ebert}, J. and {Efthymiopoulos}, I. and {Egorov}, O. and {Elsener}, K. and {Enikeev}, R. and {Ereditato}, A. and {Esposito}, L.~S. and {Fanin}, C. and {Favier}, J. and {Felici}, G. and {Ferber}, T. and {Fini}, R. and {Fournier}, L. and {Franceschi}, A. and {Frekers}, D. and {Fukuda}, T. and {Fukushima}, C. and {Galkin}, V.~I. and {Galkin}, V.~A. and {Gallet}, R. and {Gardien}, S. and {Garfagnini}, A. and {Gaudiot}, G. and {Giacomelli}, G. and {Giorgini}, M. and {Girerd}, C. and {Goellnitz}, C. and {Goeltzenlichter}, T. and {Goldberg}, J. and {Golubkov}, D. and {Gornushkin}, Y. and {Grapton}, J. -N. and {Grella}, G. and {Grianti}, F. and {Gschwendtner}, E. and {Guerin}, C. and {Guler}, M. and {Gustavino}, C. and {Guyonnet}, J. -L. and {Hagner}, C. and {Hamane}, T. and {Hara}, T. and {Hauger}, M. and {Hess}, M. and {Hierholzer}, M. and {Hoshino}, K. and {Ieva}, M. and {Incurvati}, M. and {Jakovcic}, K. and {Janicsko Csathy}, J. and {Janutta}, B. and {Jollet}, C. and {Juget}, F. and {Kazuyama}, M. and {Kim}, S.~H. and {Khovansky}, N. and {Kimura}, M. and {Klicek}, B. and {Knuesel}, J. and {Kodama}, K. and {Kolev}, D. and {Komatsu}, M. and {Kose}, U. and {Krasnoperov}, A. and {Kreslo}, I. and {Krumstein}, Z. and {Kutsenov}, V.~V. and {Kuznetsov}, V.~A. and {Laktineh}, I. and {Lavy}, M. and {Lazzaro}, C. and {Le}, T.~D. and {LeFlour}, T. and {Lenkeit}, J. and {Lewis}, J. and {Lieunard}, S. and {Ljubicic}, A. and {Longhin}, A. and {Lutter}, G. and {Malgin}, A. and {Manai}, K. and {Mandrioli}, G. and {Marotta}, A. and {Marteau}, J. and {Martin-Chassard}, G. and {Matveev}, V. and {Mauri}, N. and {Meddahi}, M. and {Meisel}, F. and {Meregaglia}, A. and {Meschini}, A. and {Messina}, M. and {Migliozzi}, P. and {Monacelli}, P. and {Monteiro}, I. and {Moreau}, F. and {Morishima}, K. and {Moser}, U. and {Muciaccia}, M.~T. and {Mugnier}, P. and {Naganawa}, N. and {Nakamura}, M. and {Nakano}, T. and {Napolitano}, T. and {Nikitina}, V. and {Niwa}, K. and {Nonoyama}, Y. and {Nozdrin}, A. and {Ogawa}, S. and {Olchevski}, A. and {Orlandi}, D. and {Orlova}, G. and {Osedlo}, V. and {Ossetski}, D. and {Paniccia}, M. and {Paoloni}, A. and {Park}, B.~D. and {Park}, I.~G. and {Pastore}, A. and {Patrizii}, L. and {Pellegrino}, L. and {Pennacchio}, E. and {Pessard}, H. and {Pilipenko}, V. and {Pistillo}, C. and {Polukhina}, N. and {Pozzato}, M. and {Pretzl}, K. and {Publichenko}, P. and {Pupilli}, F. and {Raux}, L. and {Repellin}, J.~P. and {Rescigno}, R. and {Rizhikov}, D. and {Roganova}, T. and {Romano}, G. and {Rosa}, G. and {Rostovtseva}, I. and {Rubbia}, A. and {Russo}, A. and {Ryasny}, V. and {Ryazhskaya}, O. and {Sadovski}, A. and {Sanelli}, C. and {Sato}, O. and {Sato}, Y. and {Saveliev}, V. and {Sazhina}, G. and {Schembri}, A. and {Schmidt Parzefall}, W. and {Schroeder}, H. and {Sch{\"u}tz}, H.~U. and {Schuler}, J. and {Scotto Lavina}, L. and {Serrano}, J. and {Shibuya}, H. and {Simone}, S. and {Sioli}, M. and {Sirignano}, C. and {Sirri}, G. and {Song}, J.~S. and {Spinetti}, M. and {Stanco}, L. and {Starkov}, N. and {Stipcevic}, M. and {Strauss}, T. and {Strolin}, P. and {Sugonyaev}, V. and {Takahashi}, S. and {Talochkin}, V. and {Tenti}, M. and {Tereschenko}, V. and {Terranova}, F. and {Tezuka}, I. and {Tioukov}, V. and {Tolun}, P. and {Tsarev}, V. and {Tsenov}, R. and {Tufanli}, S. and {Ugolino}, U. and {Ushida}, N. and {Van Beek}, G. and {Verguilov}, V. and {Viant}, T. and {Vilain}, P. and {Vladimirov}, M. and {Votano}, L. and {Vuilleumier}, J.~L. and {Waelchli}, T. and {Weber}, M. and {Wilquet}, G. and {Wonsak}, B. and {Wurtz}, J. and {Yakushev}, V. and {Yoon}, C.~S. and {Zaitsev}, Y. and {Zghiche}, A. and {Zimmermann}, R.},
        title = "{The OPERA experiment in the CERN to Gran Sasso neutrino beam}",
      journal = {Journal of Instrumentation},
         year = 2009,
        month = apr,
       volume = {04},
       number = {4},
        pages = {04018},
          doi = {10.1088/1748-0221/4/04/P04018},
       adsurl = {https://ui.adsabs.harvard.edu/abs/2009JInst...4.4018A}
}

@article{king2013neutrino,
  author = {{King}, Stephen F. and {Luhn}, Christoph},
        title = "{Neutrino mass and mixing with discrete symmetry}",
      journal = {Reports on Progress in Physics},
         year = 2013,
        month = may,
       volume = {76},
       number = {5},
          eid = {056201},
        pages = {056201},
          doi = {10.1088/0034-4885/76/5/056201},
archivePrefix = {arXiv},
       eprint = {1301.1340},
 primaryClass = {hep-ph},
       adsurl = {https://ui.adsabs.harvard.edu/abs/2013RPPh...76e6201K}
}

@article{vien2016delta,
  author = {{Vien}, V.~V. and {C{\'a}rcamo Hern{\'a}ndez}, A.~E. and {Long}, H.~N.},
        title = "{The {\ensuremath{\Delta}}(27) flavor 3-3-1 model with neutral leptons}",
      journal = {Nuclear Physics B},
         year = 2016,
        month = dec,
       volume = {913},
        pages = {792-814},
          doi = {10.1016/j.nuclphysb.2016.10.010},
archivePrefix = {arXiv},
       eprint = {1601.03300},
 primaryClass = {hep-ph},
       adsurl = {https://ui.adsabs.harvard.edu/abs/2016NuPhB.913..792V}
}

@article{katrin2022direct,
  author = {{Katrin Collaboration}, Aker, M. and {Beglarian}, A. and {Behrens}, J. and {Berlev}, A. and {Besserer}, U. and {Bieringer}, B. and {Block}, F. and {Bobien}, S. and {B{\"o}ttcher}, M. and {Bornschein}, B. and {Bornschein}, L. and {Brunst}, T. and {Caldwell}, T.~S. and {Carney}, R.~M.~D. and {La Cascio}, L. and {Chilingaryan}, S. and {Choi}, W. and {Debowski}, K. and {Deffert}, M. and {Descher}, M. and {D{\'\i}az Barrero}, D. and {Doe}, P.~J. and {Dragoun}, O. and {Drexlin}, G. and {Eitel}, K. and {Ellinger}, E. and {Engel}, R. and {Enomoto}, S. and {Felden}, A. and {Formaggio}, J.~A. and {Fr{\"a}nkle}, F.~M. and {Franklin}, G.~B. and {Friedel}, F. and {Fulst}, A. and {Gauda}, K. and {Gil}, W. and {Gl{\"u}ck}, F. and {Gr{\"o}ssle}, R. and {Gumbsheimer}, R. and {Gupta}, V. and {H{\"o}hn}, T. and {Hannen}, V. and {Hau{\ss}mann}, N. and {Helbing}, K. and {Hickford}, S. and {Hiller}, R. and {Hillesheimer}, D. and {Hinz}, D. and {Houdy}, T. and {Huber}, A. and {Jansen}, A. and {Karl}, C. and {Kellerer}, F. and {Kellerer}, J. and {Kleifges}, M. and {Klein}, M. and {K{\"o}hler}, C. and {K{\"o}llenberger}, L. and {Kopmann}, A. and {Korzeczek}, M. and {Koval{\'\i}k}, A. and {Krasch}, B. and {Krause}, H. and {Kunka}, N. and {Lasserre}, T. and {Le}, T.~L. and {Lebeda}, O. and {Lehnert}, B. and {Lokhov}, A. and {Machatschek}, M. and {Malcherek}, E. and {Mark}, M. and {Marsteller}, A. and {Martin}, E.~L. and {Melzer}, C. and {Menshikov}, A. and {Mertens}, S. and {Mostafa}, J. and {M{\"u}ller}, K. and {Neumann}, H. and {Niemes}, S. and {Oelpmann}, P. and {Parno}, D.~S. and {Poon}, A.~W.~P. and {Poyato}, J.~M.~L. and {Priester}, F. and {Ramachandran}, S. and {Robertson}, R.~G.~H. and {Rodejohann}, W. and {R{\"o}llig}, M. and {R{\"o}ttele}, C. and {Rodenbeck}, C. and {Ry{\v{s}}av{\'y}}, M. and {Sack}, R. and {Saenz}, A. and {Sch{\"a}fer}, P. and {Schaller N{\'e}e Pollithy}, A. and {Schimpf}, L. and {Schl{\"o}sser}, K. and {Schl{\"o}sser}, M. and {Schl{\"u}ter}, L. and {Schneidewind}, S. and {Schrank}, M. and {Schulz}, B. and {Schwemmer}, A. and {{\v{S}}ef{\v{c}}{\'\i}k}, M. and {Sibille}, V. and {Siegmann}, D. and {Slez{\'a}k}, M. and {Spanier}, F. and {Steidl}, M. and {Sturm}, M. and {Sun}, M. and {Tcherniakhovski}, D. and {Telle}, H.~H. and {Thorne}, L.~A. and {Th{\"u}mmler}, T. and {Titov}, N. and {Tkachev}, I. and {Urban}, K. and {Valerius}, K. and {V{\'e}nos}, D. and {Vizcaya Hern{\'a}ndez}, A.~P. and {Weinheimer}, C. and {Welte}, S. and {Wendel}, J. and {Wilkerson}, J.~F. and {Wolf}, J. and {W{\"u}stling}, S. and {Wydra}, J. and {Xu}, W. and {Yen}, Y. -R. and {Zadoroghny}, S. and {Zeller}, G.},
        title = "{Direct neutrino-mass measurement with sub-electronvolt sensitivity}",
      journal = {Nature Physics},
         year = 2022,
        month = feb,
       volume = {18},
       number = {2},
        pages = {160-166},
          doi = {10.1038/s41567-021-01463-1},
       adsurl = {https://ui.adsabs.harvard.edu/abs/2022NatPh..18..160K}
}

@article{particle2022review,
  author = {{Workman}, R.~L. and {Burkert}, V.~D. and {Crede}, V. and {Klempt}, E. and {Thoma}, U. and {Tiator}, L. and {Agashe}, K. and {Aielli}, G. and {Allanach}, B.~C. and {Amsler}, C. and {Antonelli}, M. and {Aschenauer}, E.~C. and {Asner}, D.~M. and {Baer}, H. and {Banerjee}, Sw and {Barnett}, R.~M. and {Baudis}, L. and {Bauer}, C.~W. and {Beatty}, J.~J. and {Belousov}, V.~I. and {Beringer}, J. and {Bettini}, A. and {Biebel}, O. and {Black}, K.~M. and {Blucher}, E. and {Bonventre}, R. and {Bryzgalov}, V.~V. and {Buchmuller}, O. and {Bychkov}, M.~A. and {Cahn}, R.~N. and {Carena}, M. and {Ceccucci}, A. and {Cerri}, A. and {Chivukula}, R. Sekhar and {Cowan}, G. and {Cranmer}, K. and {Cremonesi}, O. and {D'Ambrosio}, G. and {Damour}, T. and {de Florian}, D. and {de Gouv{\^e}a}, A. and {DeGrand}, T. and {de Jong}, P. and {Demers}, S. and {Dobrescu}, B.~A. and {D'Onofrio}, M. and {Doser}, M. and {Dreiner}, H.~K. and {Eerola}, P. and {Egede}, U. and {Eidelman}, S. and {El-Khadra}, A.~X. and {Ellis}, J. and {Eno}, S.~C. and {Erler}, J. and {Ezhela}, V.~V. and {Fetscher}, W. and {Fields}, B.~D. and {Freitas}, A. and {Gallagher}, H. and {Gershtein}, Y. and {Gherghetta}, T. and {Gonzalez-Garcia}, M.~C. and {Goodman}, M. and {Grab}, C. and {Gritsan}, A.~V. and {Grojean}, C. and {Groom}, D.~E. and {Gr{\"u}newald}, M. and {Gurtu}, A. and {Gutsche}, T. and {Haber}, H.~E. and {Hamel}, Matthieu and {Hanhart}, C. and {Hashimoto}, S. and {Hayato}, Y. and {Hebecker}, A. and {Heinemeyer}, S. and {Hern{\'a}ndez-Rey}, J.~J. and {Hikasa}, K. and {Hisano}, J. and {H{\"o}cker}, A. and {Holder}, J. and {Hsu}, L. and {Huston}, J. and {Hyodo}, T. and {Ianni}, Al and {Kado}, M. and {Karliner}, M. and {Katz}, U.~F. and {Kenzie}, M. and {Khoze}, V.~A. and {Klein}, S.~R. and {Krauss}, F. and {Kreps}, M. and {Kri{\v{z}}an}, P. and {Krusche}, B. and {Kwon}, Y. and {Lahav}, O. and {Laiho}, J. and {Lellouch}, L.~P. and {Lesgourgues}, J. and {Liddle}, A.~R. and {Ligeti}, Z. and {Lin}, C. -J. and {Lippmann}, C. and {Liss}, T.~M. and {Littenberg}, L. and {Louren{\c{c}}o}, C. and {Lugovsky}, K.~S. and {Lugovsky}, S.~B. and {Lusiani}, A. and {Makida}, Y. and {Maltoni}, F. and {Mannel}, T. and {Manohar}, A.~V. and {Marciano}, W.~J. and {Masoni}, A. and {Matthews}, J. and {Mei{\ss}ner}, U. -G. and {Melzer-Pellmann}, I. -A. and {Mikhasenko}, M. and {Miller}, D.~J. and {Milstead}, D. and {Mitchell}, R.~E. and {M{\"o}nig}, K. and {Molaro}, P. and {Moortgat}, F. and {Moskovic}, M. and {Nakamura}, K. and {Narain}, M. and {Nason}, P. and {Navas}, S. and {Nelles}, A. and {Neubert}, M. and {Nevski}, P. and {Nir}, Y. and {Olive}, K.~A. and {Patrignani}, C. and {Peacock}, J.~A. and {Petrov}, V.~A. and {Pianori}, E. and {Pich}, A. and {Piepke}, A. and {Pietropaolo}, F. and {Pomarol}, A. and {Pordes}, S. and {Profumo}, S. and {Quadt}, A. and {Rabbertz}, K. and {Rademacker}, J. and {Raffelt}, G. and {Ramsey-Musolf}, M. and {Ratcliff}, B.~N. and {Richardson}, P. and {Ringwald}, A. and {Robinson}, D.~J. and {Roesler}, S. and {Rolli}, S. and {Romaniouk}, A. and {Rosenberg}, L.~J. and {Rosner}, J.~L. and {Rybka}, G. and {Ryskin}, M.~G. and {Ryutin}, R.~A. and {Sakai}, Y. and {Sarkar}, S. and {Sauli}, F. and {Schneider}, O. and {Sch{\"o}nert}, S. and {Scholberg}, K. and {Schwartz}, A.~J. and {Schwiening}, J. and {Scott}, D. and {Sefkow}, F. and {Seljak}, U. and {Sharma}, V. and {Sharpe}, S.~R. and {Shiltsev}, V. and {Signorelli}, G. and {Silari}, M. and {Simon}, F. and {Sj{\"o}strand}, T. and {Skands}, P. and {Skwarnicki}, T. and {Smoot}, G.~F. and {Soffer}, A. and {Sozzi}, M.~S. and {Spanier}, S. and {Spiering}, C. and {Stahl}, A. and {Stone}, S.~L. and {Sumino}, Y. and {Syphers}, M.~J. and {Takahashi}, F. and {Tanabashi}, M. and {Tanaka}, J. and {Ta{\v{s}}evsk{\'y}}, M. and {Terao}, K. and {Terashi}, K. and {Terning}, J. and {Thorne}, R.~S. and {Titov}, M. and {Tkachenko}, N.~P. and {Tovey}, D.~R. and {Trabelsi}, K. and {Urquijo}, P. and {Valencia}, G. and {Van de Water}, R. and {Varelas}, N. and {Venanzoni}, G. and {Verde}, L. and {Vivarelli}, I. and {Vogel}, P. and {Vogelsang}, W. and {Vorobyev}, V. and {Wakely}, S.~P. and {Walkowiak}, W. and {Walter}, C.~W. and {Wands}, D. and {Weinberg}, D.~H. and {Weinberg}, E.~J. and {Wermes}, N. and {White}, M. and {Wiencke}, L.~R. and {Willocq}, S. and {Wohl}, C.~G. and {Woody}, C.~L. and {Yao}, W. -M. and {Yokoyama}, M. and {Yoshida}, R. and {Zanderighi}, G. and {Zeller}, G.~P. and {Zenin}, O.~V. and {Zhu}, R. -Y. and {Zhu}, Shi-Lin and {Zimmermann}, F. and {Zyla}, P.~A. and {Particle Data Group}},
        title = "{Review of Particle Physics}",
      journal = {Progress of Theoretical and Experimental Physics},
         year = 2022,
        month = aug,
       volume = {2022},
       number = {8},
          eid = {083C01},
        pages = {083C01},
          doi = {10.1093/ptep/ptac097},
       adsurl = {https://ui.adsabs.harvard.edu/abs/2022PTEP.2022h3C01W}
}

@article{hirsch2009a4,
  author = {{Hirsch}, M. and {Morisi}, S. and {Valle}, J.~W.~F.},
        title = "{A$_4$-based tri-bimaximal mixing within inverse and linear seesaw schemes}",
      journal = {Physics Letters B},
         year = 2009,
        month = sep,
       volume = {679},
       number = {5},
        pages = {454-459},
          doi = {10.1016/j.physletb.2009.08.003},
archivePrefix = {arXiv},
       eprint = {0905.3056},
 primaryClass = {hep-ph},
       adsurl = {https://ui.adsabs.harvard.edu/abs/2009PhLB..679..454H}
}

@article{carcamo2021controlled,
  author = {{C{\'a}rcamo Hern{\'a}ndez}, A.~E. and {de Medeiros Varzielas}, Ivo and {L{\'o}pez-Ib{\'a}{\~n}ez}, M.~L. and {Melis}, Aurora},
        title = "{Controlled fermion mixing and FCNCs in a {\ensuremath{\Delta}}(27) 3+1 Higgs Doublet Model}",
      journal = {Journal of High Energy Physics},
         year = 2021,
        month = may,
       volume = {2021},
       number = {5},
          eid = {215},
        pages = {215},
          doi = {10.1007/JHEP05(2021)215},
archivePrefix = {arXiv},
       eprint = {2102.05658},
 primaryClass = {hep-ph},
       adsurl = {https://ui.adsabs.harvard.edu/abs/2021JHEP...05..215C}
}

@article{sruthilaya2018a_4,
  author = {{Sruthilaya}, M. and {Mohanta}, Rukmani and {Patra}, Sudhanwa},
        title = "{A$_4$ realization of linear seesaw and neutrino phenomenology}",
      journal = {European Physical Journal C},
         year = 2018,
        month = sep,
       volume = {78},
       number = {9},
          eid = {719},
        pages = {719},
          doi = {10.1140/epjc/s10052-018-6181-6},
archivePrefix = {arXiv},
       eprint = {1709.01737},
 primaryClass = {hep-ph},
       adsurl = {https://ui.adsabs.harvard.edu/abs/2018EPJC...78..719S}
}

@article{borah2019linear,
  author = {{Borah}, Debasish and {Karmakar}, Biswajit},
        title = "{Linear seesaw for Dirac neutrinos with A$_{4}$ flavour symmetry}",
      journal = {Physics Letters B},
         year = 2019,
        month = feb,
       volume = {789},
        pages = {59-70},
          doi = {10.1016/j.physletb.2018.12.006},
archivePrefix = {arXiv},
       eprint = {1806.10685},
 primaryClass = {hep-ph},
       adsurl = {https://ui.adsabs.harvard.edu/abs/2019PhLB..789...59B}
}

@article{T71,
  author = {{Luhn}, Christoph and {Nasri}, Salah and {Ramond}, Pierre},
        title = "{Tri-bimaximal neutrino mixing and the family symmetry Z$_{7}${\ensuremath{\rtimes}}Z$_{3}$}",
      journal = {Physics Letters B},
         year = 2007,
        month = aug,
       volume = {652},
       number = {1},
        pages = {27-33},
          doi = {10.1016/j.physletb.2007.06.059},
archivePrefix = {arXiv},
       eprint = {0706.2341},
 primaryClass = {hep-ph},
       adsurl = {https://ui.adsabs.harvard.edu/abs/2007PhLB..652...27L}
}

@article{T72,
  author = {{Hagedorn}, Claudia and {Schmidt}, Michael A. and {Smirnov}, Alexei Yu.},
        title = "{Lepton mixing and cancellation of the Dirac mass hierarchy in SO(10) GUTs with flavor symmetries T$_{7}$ and {\ensuremath{\Sigma}}(81)}",
      journal = {Physical Review D},
         year = 2009,
        month = feb,
       volume = {79},
       number = {3},
          eid = {036002},
        pages = {036002},
          doi = {10.1103/PhysRevD.79.036002},
archivePrefix = {arXiv},
       eprint = {0811.2955},
 primaryClass = {hep-ph},
       adsurl = {https://ui.adsabs.harvard.edu/abs/2009PhRvD..79c6002H}
}

@article{T73,
  author = {{Cao}, Qing-Hong and {Khalil}, Shaaban and {Ma}, Ernest and {Okada}, Hiroshi},
        title = "{Observable T$_{7}$ Lepton Flavor Symmetry at the Large Hadron Collider}",
      journal = {Physical Review Letters},
         year = 2011,
        month = apr,
       volume = {106},
       number = {13},
          eid = {131801},
        pages = {131801},
          doi = {10.1103/PhysRevLett.106.131801},
archivePrefix = {arXiv},
       eprint = {1009.5415},
 primaryClass = {hep-ph},
       adsurl = {https://ui.adsabs.harvard.edu/abs/2011PhRvL.106m1801C}
}

@article{T74,
  author = {{Luhn}, Christoph and {Parattu}, Krishna Mohan and {Wingerter}, Ak{\i}n},
        title = "{A minimal model of neutrino flavor}",
      journal = {Journal of High Energy Physics},
         year = 2012,
        month = dec,
       volume = {2012},
          eid = {96},
        pages = {96},
          doi = {10.1007/JHEP12(2012)096},
archivePrefix = {arXiv},
       eprint = {1210.1197},
 primaryClass = {hep-ph},
       adsurl = {https://ui.adsabs.harvard.edu/abs/2012JHEP...12..096L}
}

@article{T75,
  author = {{Kajiyama}, Yuji and {Okada}, Hiroshi and {Yagyu}, Kei},
        title = "{T $_{7}$ flavor model in three loop seesaw and Higgs phenomenology}",
      journal = {Journal of High Energy Physics},
         year = 2013,
        month = oct,
       volume = {2013},
          eid = {196},
        pages = {196},
          doi = {10.1007/JHEP10(2013)196},
archivePrefix = {arXiv},
       eprint = {1307.0480},
 primaryClass = {hep-ph},
       adsurl = {https://ui.adsabs.harvard.edu/abs/2013JHEP...10..196K}
}

@article{T76,
  author = {{Bonilla}, Cesar and {Morisi}, Stefano and {Peinado}, Eduardo and {Valle}, J.~W.~F.},
        title = "{Relating quarks and leptons with the T$_{7}$ flavour group}",
      journal = {Physics Letters B},
         year = 2015,
        month = mar,
       volume = {742},
        pages = {99-106},
          doi = {10.1016/j.physletb.2015.01.017},
archivePrefix = {arXiv},
       eprint = {1411.4883},
 primaryClass = {hep-ph},
       adsurl = {https://ui.adsabs.harvard.edu/abs/2015PhLB..742...99B}
}

@article{T77,
  author = {{Vien}, V.~V. and {Long}, H.~N.},
        title = "{The T $_{7}$ flavor symmetry in 3-3-1 model with neutral leptons}",
      journal = {Journal of High Energy Physics},
         year = 2014,
        month = apr,
       volume = {2014},
          eid = {133},
        pages = {133},
          doi = {10.1007/JHEP04(2014)133},
archivePrefix = {arXiv},
       eprint = {1402.1256},
 primaryClass = {hep-ph},
       adsurl = {https://ui.adsabs.harvard.edu/abs/2014JHEP...04..133V}
}

@article{T78,
  author = {{Vien}, V.~V.},
        title = "{T$_{7}$ flavor symmetry scheme for understanding neutrino mass and mixing in 3-3-1 model with neutral leptons}",
      journal = {Modern Physics Letters A},
         year = 2014,
        month = sep,
       volume = {29},
       number = {28},
          eid = {1450139},
        pages = {1450139},
          doi = {10.1142/S0217732314501399},
archivePrefix = {arXiv},
       eprint = {1508.02585},
 primaryClass = {hep-ph},
       adsurl = {https://ui.adsabs.harvard.edu/abs/2014MPLA...2950139V}
}

@article{T79,
  author = {{C{\'a}rcamo Hern{\'a}ndez}, A.~E. and {Martinez}, R.},
        title = "{Fermion mass and mixing pattern in a minimal T$_7$ flavor 331 model}",
      journal = {Journal of Physics G: Nuclear and Particle Physics},
         year = 2016,
        month = feb,
     volume = {43},
       number = {4},
        pages = {45003},
          doi = {10.1088/0954-3899/43/4/045003},
archivePrefix = {arXiv},
       eprint = {1511.07997},
 primaryClass = {hep-ph},
       adsurl = {https://ui.adsabs.harvard.edu/abs/2015arXiv151107997C}
}

@article{T710,
  author = {{Arbel{\'a}ez}, Carolina and {C{\'a}rcamo Hern{\'a}ndez}, A.~E. and {Kovalenko}, Sergey and {Schmidt}, Iv{\'a}n},
        title = "{Adjoint S U (5 ) GUT model with T$_{7}$ flavor symmetry}",
      journal = {Physical Review D},
         year = 2015,
        month = dec,
       volume = {92},
       number = {11},
          eid = {115015},
        pages = {115015},
          doi = {10.1103/PhysRevD.92.115015},
archivePrefix = {arXiv},
       eprint = {1507.03852},
 primaryClass = {hep-ph},
       adsurl = {https://ui.adsabs.harvard.edu/abs/2015PhRvD..92k5015A}
}

@article{T713,
  author = {{Vien}, V.~V. and {Long}, H.~N.},
        title = "{Fermion Mass and Mixing in a Simple Extension of the Standard Model Based on T $_{7}$ Flavor Symmetry}",
      journal = {Physics of Atomic Nuclei},
         year = 2019,
        month = mar,
       volume = {82},
       number = {2},
        pages = {168-182},
          doi = {10.1134/S1063778819020133},
       adsurl = {https://ui.adsabs.harvard.edu/abs/2019PAN....82..168V}
}

@book{bookfordiscretesymmetries,
author = {{Ishimori}, Hajime and {Kobayashi}, Tatsuo and {Ohki}, Hiroshi and {Okada}, Hiroshi and {Shimizu}, Yusuke and {Tanimoto}, Morimitsu},
        title = "{An Introduction to Non-Abelian Discrete Symmetries for Particle Physicists}",
         year = 2012,
       volume = {858},
          doi = {10.1007/978-3-642-30805-5},
       adsurl = {https://ui.adsabs.harvard.edu/abs/2012LNP...858.....I},
publisher={Springer}
}

@article{panda2008comparison,
  title="{Comparison of particle swarm optimization and genetic algorithm for FACTS-based controller design}",
  author={Panda, Sidhartha and Padhy, Narayana Prasad},
  journal={Applied soft computing},
  volume={8},
  number={4},
  pages={1418--1427},
  year={2008},
  publisher={Elsevier}
}

@article{okwu2021particle,
  title="{Particle swarm optimisation}",
  author={Okwu, Modestus O and Tartibu, Lagouge K and Okwu, Modestus O and Tartibu, Lagouge K},
  journal={Metaheuristic Optimization: Nature-Inspired Algorithms Swarm and Computational Intelligence, Theory and Applications},
  pages={5--13},
  year={2021},
  publisher={Springer}
}

@article{yu2020comparison,
  title="{A Comparison between Generic Algorithm and Particle Swarm Optimization}",
  author={Yu, YongHai and Yin, ShuE},
  booktitle={Proceedings of the 1st International Symposium on Artificial Intelligence in Medical Sciences},
  pages={137--139},
  year={2020}
}

@article{stacey2003particle,
  title="{Particle swarm optimization with mutation}",
  author={Stacey, Andrew and Jancic, Mirjana and Grundy, Ian},
  booktitle={The 2003 Congress on Evolutionary Computation, 2003. CEC'03.},
  volume={2},
  pages={1425--1430},
  year={2003},
  organization={IEEE}
}

@article{eberhart1995new,
  title="{A new optimizer using particle swarm theory}",
  author={Eberhart, Russell and Kennedy, James},
  booktitle={MHS'95. Proceedings of the sixth international symposium on micro machine and human science},
  pages={39--43},
  year={1995},
  organization={Ieee}
}

@article{toushmalani2013gravity,
  title="{Gravity inversion of a fault by Particle swarm optimization (PSO)}",
  author={Toushmalani, Reza},
  journal={SpringerPlus},
  volume={2},
  number={1},
  pages={1--7},
  year={2013},
  publisher={SpringerOpen}
}

@article{bassi2011automatic,
  title="{Automatic tuning of proportional-integral-derivative (PID) controller using particle swarm optimization (PSO) algorithm}",
  author={Bassi, SJ and Mishra, MK and Omizegba, EE},
  journal={International Journal of Artificial Intelligence \& Applications},
  volume={2},
  number={4},
  pages={25},
  year={2011},
  publisher={Academy \& Industry Research Collaboration Center (AIRCC)}
}

@article{esmin2015review,
  title="{A review on particle swarm optimization algorithm and its variants to clustering high-dimensional data}",
  author={Esmin, Ahmed AA and Coelho, Rodrigo A and Matwin, Stan},
  journal={Artificial Intelligence Review},
  volume={44},
  pages={23--45},
  year={2015},
  publisher={Springer}
}

@article{rana2011review,
  title="{A review on particle swarm optimization algorithms and their applications to data clustering}",
  author={Rana, Sandeep and Jasola, Sanjay and Kumar, Rajesh},
  journal={Artificial Intelligence Review},
  volume={35},
  pages={211--222},
  year={2011},
  publisher={Springer}
}

@article{he2007parameter,
  title="{Parameter estimation for chaotic systems by particle swarm optimization}",
  author={He, Qie and Wang, Ling and Liu, Bo},
  journal={Chaos, Solitons \& Fractals},
  volume={34},
  number={2},
  pages={654--661},
  year={2007},
  publisher={Elsevier}
}

@article{alatas2009chaos,
  title="{Chaos embedded particle swarm optimization algorithms}",
  author={Alatas, Bilal and Akin, Erhan and Ozer, A Bedri},
  journal={Chaos, Solitons \& Fractals},
  volume={40},
  number={4},
  pages={1715--1734},
  year={2009},
  publisher={Elsevier}
}

@article{babazadeh2009optimization,
  title="{Optimization of fuel core loading pattern design in a VVER nuclear power reactors using Particle Swarm Optimization (PSO)}",
  author={Babazadeh, Davood and Boroushaki, Mehrdad and Lucas, Caro},
  journal={Annals of nuclear energy},
  volume={36},
  number={7},
  pages={923--930},
  year={2009},
  publisher={Elsevier}
}

@article{ibrahim2019hybridization,
  title="{A hybridization of cuckoo search and particle swarm optimization for solving nonlinear systems}",
  author={Ibrahim, Abdelmonem M and Tawhid, Mohamed A},
  journal={Evolutionary Intelligence},
  volume={12},
  pages={541--561},
  year={2019},
  publisher={Springer}
}

@article{subbaraj2010hybrid,
  title="{Hybrid particle swarm optimization based optimal reactive power dispatch}",
  author={Subbaraj, P and Rajnarayanan, PN},
  journal={International Journal of Computer Applications},
  volume={1},
  number={5},
  pages={65--70},
  year={2010},
  publisher={Citeseer}
}

@article{jiang2020multilayer,
  title="{Multilayer optical thin film design with deep Q learning}",
  author={Jiang, Anqing and Osamu, Yoshie and Chen, Liangyao},
  journal={Scientific reports},
  volume={10},
  number={1},
  pages={12780},
  year={2020},
  publisher={Nature Publishing Group UK London}
}

@article{yue2019determination,
  title="{Determination of thin metal film’s thickness and optical constants based on SPR phase detection by simulated annealing particle swarm optimization}",
  author={Yue, Chong and Qin, Zirui and Lang, Yaopu and Liu, Qinggang},
  journal={Optics Communications},
  volume={430},
  pages={238--245},
  year={2019},
  publisher={Elsevier}
}

@article{rabady2014global,
  title="{Global optimal design of optical multilayer thin-film filters using particle swarm optimization}",
  author={Rabady, Rabi I and Ababneh, Almahdi},
  journal={Optik},
  volume={125},
  number={1},
  pages={548--553},
  year={2014},
  publisher={Elsevier}
}

@article{ruan2016determination,
  title="{Determination of optical properties and thickness of optical thin film using stochastic particle swarm optimization}",
  author={Ruan, Zhao-Hui and Yuan, Yuan and Zhang, Xiao-Xian and Shuai, Yong and Tan, He-Ping},
  journal={Solar Energy},
  volume={127},
  pages={147--158},
  year={2016},
  publisher={Elsevier}
}

@article{mehmood2019nature,
  title="{Nature-inspired heuristic paradigms for parameter estimation of control autoregressive moving average systems}",
  author={Mehmood, Ammara and Zameer, Aneela and Raja, Muhammad Asif Zahoor and Bibi, Rabia and Chaudhary, Naveed Ishtiaq and Aslam, Muhammad Saeed},
  journal={Neural Computing and Applications},
  volume={31},
  pages={5819--5842},
  year={2019},
  publisher={Springer}
}

@article{yetis2014forecasting,
  title="{Forecasting of Turkey's electricity consumption using Artificial Neural Network}",
  author={Yetis, Yunus and Jamshidi, Mo},
  booktitle={2014 World Automation Congress (WAC)},
  pages={723--728},
  year={2014},
  organization={IEEE}
}

@article{yassin2016binary,
  title="{Binary particle swarm optimization structure selection of nonlinear autoregressive moving average with exogenous inputs (NARMAX) model of a flexible robot arm}",
  author={Yassin, Ihsan Mohd and Zabidi, Azlee and Amin Megat Ali, Megat Syahirul and Md Tahir, Nooritawati and Zainol Abidin, Husna and Rizman, Zairi Ismael},
  journal={International Journal on Advanced Science, Engineering and Information Technology},
  volume={6},
  number={5},
  pages={630--637},
  year={2016}
}

@article{akbar2019novel,
  title="{Novel application of FO-DPSO for 2-D parameter estimation of electromagnetic plane waves}",
  author={Akbar, Sadiq and Zaman, Fawad and Asif, Muhammad and Rehman, Ata Ur and Raja, Muhammad Asif Zahoor},
  journal={Neural Computing and Applications},
  volume={31},
  pages={3681--3690},
  year={2019},
  publisher={Springer}
}

@article{cowan1956detection,
  author = "Cowan, C. L. and Reines, F. and Harrison, F. B. and Kruse, H. W. and McGuire, A. D.",
    title = "{Detection of the free neutrino: A Confirmation}",
    doi = "10.1126/science.124.3212.103",
    journal = "Science",
    volume = "124",
    pages = "103--104",
    year = "1956"
}

@article{raffelt2008neutrinos,
  title={Neutrinos and Stars},
  author={Raffelt, Georg G},
  booktitle={Neutrinos in Particle Physics, Astrophysics and Cosmology},
  pages={185--194},
  year={2008},
archivePrefix = {arXiv},
       eprint = {1201.1637},
 primaryClass = {astro-ph.SR},
       adsurl = {https://ui.adsabs.harvard.edu/abs/2012arXiv1201.1637R}
}

@article{burrows1990neutrinos,
  author = "Burrows, A.",
    title = "{Neutrinos From Supernova Explosions}",
    reportNumber = "STEWARD-925, ATAP-90-09",
    doi = "10.1146/annurev.ns.40.120190.001145",
    journal = "Ann. Rev. Nucl. Part. Sci.",
    volume = "40",
    pages = "181--212",
    year = "1990"
}

@article{tamborra2018neutrinos,
  author = "Tamborra, Irene and Murase, Kohta",
    title = "{Neutrinos from Supernovae}",
    doi = "10.1007/s11214-018-0468-7",
    journal = "Space Sci. Rev.",
    volume = "214",
    number = "1",
    pages = "31",
    year = "2018"
}

@article{burrows2000neutrinos,
  author = "Burrows, A. and Young, T.",
    title = "{Neutrinos and supernova theory}",
    doi = "10.1016/S0370-1573(00)00016-8",
    journal = "Phys. Rept.",
    volume = "333",
    pages = "63--75",
    year = "2000"
}

@article{cooperstein1988neutrinos,
  author = "Cooperstein, J.",
    title = "{Neutrinos in Supernovae}",
    doi = "10.1016/0370-1573(88)90038-5",
    journal = "Phys. Rept.",
    volume = "163",
    pages = "95--126",
    year = "1988"
}

@article{herant1997neutrinos,
  title="{Neutrinos and supernovae}",
  author={Herant, Marc and Colgate, Stirling A and Benz, Willy and Fryer, Chris},
  journal={Los Alamos Science},
  volume={25},
  pages={64--78},
  year={1997}
}

@article{sarkar2003neutrinos,
 author = {{Sarkar}, Subir},
        title = "{Neutrinos from the Big Bang}",
      journal = {arXiv e-prints},
         year = 2003,
        month = feb,
          eid = {hep-ph/0302175},
        pages = {hep-ph/0302175},
          doi = {10.48550/arXiv.hep-ph/0302175},
archivePrefix = {arXiv},
       eprint = {hep-ph/0302175},
 primaryClass = {hep-ph},
       adsurl = {https://ui.adsabs.harvard.edu/abs/2003hep.ph....2175S}
}

@article{steigman2005neutrinos,
  author = {{Steigman}, Gary},
        title = "{Neutrinos and Big Bang Nucleosynthesis}",
      journal = {Physica Scripta Volume T},
         year = 2005,
        month = jan,
       volume = {121},
        pages = {142-146},
          doi = {10.1088/0031-8949/2005/T121/021},
archivePrefix = {arXiv},
       eprint = {hep-ph/0501100},
 primaryClass = {hep-ph},
       adsurl = {https://ui.adsabs.harvard.edu/abs/2005PhST..121..142S}
}

@article{blennow2013probing,
 author = {{Blennow}, Mattias and {Carrigan}, Marcus and {Fernandez Martinez}, Enrique},
        title = "{Probing the Dark Matter mass and nature with neutrinos}",
      journal = {Journal of Cosmology and Astroparticle Physics},
         year = 2013,
        month = jun,
       volume = {2013},
       number = {6},
          eid = {038},
        pages = {038},
          doi = {10.1088/1475-7516/2013/06/038},
archivePrefix = {arXiv},
       eprint = {1303.4530},
 primaryClass = {hep-ph},
       adsurl = {https://ui.adsabs.harvard.edu/abs/2013JCAP...06..038B}
}

@article{agarwalla2011neutrino,
  author = "Agarwalla, Sanjib Kumar and Blennow, Mattias and Fernandez Martinez, Enrique and Mena, Olga",
    title = "{Neutrino Probes of the Nature of Light Dark Matter}",
    eprint = "1105.4077",
    archivePrefix = "arXiv",
    primaryClass = "hep-ph",
    reportNumber = "CERN-PH-TH-2011-118, EURONU-WP6-11-35, IFIC-11-24, MPP-2011-56",
    doi = "10.1088/1475-7516/2011/09/004",
    journal = "JCAP",
    volume = "09",
    pages = "004",
    year = "2011"
}

@article{dodelson1994sterile,
  author = {{Dodelson}, Scott and {Widrow}, Lawrence M.},
        title = "{Sterile neutrinos as dark matter}",
      journal = {Physical Review Letters},
         year = 1994,
        month = jan,
       volume = {72},
       number = {1},
        pages = {17-20},
          doi = {10.1103/PhysRevLett.72.17},
archivePrefix = {arXiv},
       eprint = {hep-ph/9303287},
 primaryClass = {hep-ph},
       adsurl = {https://ui.adsabs.harvard.edu/abs/1994PhRvL..72...17D}
}

@article{bertone2018new,
  author = "Bertone, Gianfranco and Tait, M. P., Tim",
    title = "{A new era in the search for dark matter}",
    eprint = "1810.01668",
    archivePrefix = "arXiv",
    primaryClass = "astro-ph.CO",
    doi = "10.1038/s41586-018-0542-z",
    journal = "Nature",
    volume = "562",
    number = "7725",
    pages = "51--56",
    year = "2018"
}

@article{hooper2008strategies,
  author = "Hooper, Dan and Baltz, Edward A.",
    title = "{Strategies for Determining the Nature of Dark Matter}",
    eprint = "0802.0702",
    archivePrefix = "arXiv",
    primaryClass = "hep-ph",
    reportNumber = "SLAC-PUB-13146, FERMILAB-PUB-08-026-A",
    doi = "10.1146/annurev.nucl.58.110707.171217",
    journal = "Ann. Rev. Nucl. Part. Sci.",
    volume = "58",
    pages = "293--314",
    year = "2008"
}

@article{t2k2020constraint,
  author = "Abe, K. and others",
    collaboration = "T2K",
    title = "{Constraint on the matter\textendash{}antimatter symmetry-violating phase in neutrino oscillations}",
    eprint = "1910.03887",
    archivePrefix = "arXiv",
    primaryClass = "hep-ex",
    doi = "10.1038/s41586-020-2177-0",
    journal = "Nature",
    volume = "580",
    number = "7803",
    pages = "339--344",
    year = "2020",
    note = "[Erratum: Nature 583, E16 (2020)]"
}

@article{weinheimer2013neutrino,
  author = "Weinheimer, Christian and Zuber, Kai",
    title = "{Neutrino Masses}",
    eprint = "1307.3518",
    archivePrefix = "arXiv",
    primaryClass = "hep-ex",
    doi = "10.1002/andp.201300063",
    journal = "Annalen Phys.",
    volume = "525",
    number = "8-9",
    pages = "565--575",
    year = "2013"
}

@book{bilenky2018introduction,
  author = "Bilenky, Samoil",
    title = "{Introduction to the Physics of Massive and Mixed Neutrinos}",
    doi = "10.1007/978-3-319-74802-3",
    isbn = "978-3-319-74801-6, 978-3-319-74802-3",
    publisher = "Springer",
    volume = "947",
    year = "2018"
}

@article{chakraborty2014higgs,
  author = {{Chakraborty}, Sabyasachi and {Roy}, Sourov},
        title = "{Higgs boson mass, neutrino masses and mixing and keV dark matter in an U(1)$_{ R }$ - lepton number model}",
      journal = {Journal of High Energy Physics},
         year = 2014,
        month = jan,
       volume = {2014},
          eid = {101},
        pages = {101},
          doi = {10.1007/JHEP01(2014)101},
archivePrefix = {arXiv},
       eprint = {1309.6538},
 primaryClass = {hep-ph},
       adsurl = {https://ui.adsabs.harvard.edu/abs/2014JHEP...01..101C}
}

@article{guedria2016improved,
  title="{Improved accelerated PSO algorithm for mechanical engineering optimization problems}",
  author={Guedria, Najeh Ben},
  journal={Applied Soft Computing},
  volume={40},
  pages={455--467},
  year={2016},
  publisher={Elsevier}
}

@article{kaveh2013engineering,
  title="{Engineering design optimization using a hybrid PSO and HS algorithm}",
  author={Kaveh, A and NASR, ELAHI A},
  year={2013},
  publisher={ASIAN JOURNAL OF CIVIL ENGINEERING (BUILDING AND HOUSING)}
}

@article{kumar2021design,
  title="{Design of an efficient hybridized CS-PSO algorithm and its applications for solving constrained and bound constrained structural engineering design problems}",
  author={Kumar, Nirmal and Mahato, Sanat Kumar and Bhunia, Asoke Kumar},
  journal={Results in Control and Optimization},
  volume={5},
  pages={100064},
  year={2021},
  publisher={Elsevier}
}

@article{djemame2019solving,
  title="{Solving reverse emergence with quantum PSO application to image processing}",
  author={Djemame, Safia and Batouche, Mohamed and Oulhadj, Hamouche and Siarry, Patrick},
  journal={Soft Computing},
  volume={23},
  number={16},
  pages={6921--6935},
  year={2019},
  publisher={Springer}
}

@article{singh2014image,
  title="{Image contrast enhancement using GA and PSO: a survey}",
  author={Singh, Ravindra Pal and Dixit, Manish and Silakari, Sanjay},
  booktitle={2014 International Conference on Computational Intelligence and Communication Networks},
  pages={186--189},
  year={2014},
  organization={IEEE}
}

@article{pramanik2015image,
  title="{Image registration using PSO and APSO: a comparative analysis}",
  author={Pramanik, Jitendra and Dalai, Sunita and Rana, Debaraj},
  journal={International Journal of Computer Applications},
  volume={116},
  number={21},
  year={2015},
  publisher={Citeseer}
}

@article{azayite2019financial,
  title="{Financial Early Warning System Model Based on Neural Networks, PSO and SA Algorithms}",
  author={Azayite, Fatima Zahra and Achchab, Said},
  booktitle={EngOpt 2018 Proceedings of the 6th International Conference on Engineering Optimization},
  pages={970--982},
  year={2019},
  organization={Springer}
}

@article{chiam2009memetic,
  title="{A memetic model of evolutionary PSO for computational finance applications}",
  author={Chiam, Swee Chiang and Tan, Kay Chen and Mamun, A Al},
  journal={Expert Systems with Applications},
  volume={36},
  number={2},
  pages={3695--3711},
  year={2009},
  publisher={Elsevier}
}

@article{pan2022design,
  title="{Design of financial management model using the forward neural network based on particle swarm optimization algorithm}",
  author={Pan, Yilin and others},
  journal={Computational Intelligence and Neuroscience},
  volume={2022},
  year={2022},
  publisher={Hindawi}
}

@article{marinakis2009ant,
  title="{Ant colony and particle swarm optimization for financial classification problems}",
  author={Marinakis, Yannis and Marinaki, Magdalene and Doumpos, Michael and Zopounidis, Constantin},
  journal={Expert Systems with Applications},
  volume={36},
  number={7},
  pages={10604--10611},
  year={2009},
  publisher={Elsevier}
}

@article{meissner2006optimized,
  title="{Optimized Particle Swarm Optimization (OPSO) and its application to artificial neural network training}",
  author={Meissner, Michael and Schmuker, Michael and Schneider, Gisbert},
  journal={BMC bioinformatics},
  volume={7},
  number={1},
  pages={1--11},
  year={2006},
  publisher={BioMed Central}
}

@article{rauf2018training,
  title="{Training of artificial neural network using pso with novel initialization technique}",
  author={Rauf, Hafiz Tayyab and Bangyal, Waqas Haider and Ahmad, Jamil and Bangyal, Saad Abdullah},
  booktitle={2018 international conference on innovation and intelligence for informatics, computing, and technologies (3ICT)},
  pages={1--8},
  year={2018},
  organization={IEEE}
}

@article{Model2023,
  title="{A realistic neutrino mixing scheme arising from $A_4$ symmetry}",
  author={Dey, Manash and Roy, Subhankar},
  journal={Europhysics Letters},
  year={2023}
}

@article{cai2018lepton,
  author = "Cai, Yi and Han, Tao and Li, Tong and Ruiz, Richard",
    title = "{Lepton Number Violation: Seesaw Models and Their Collider Tests}",
    eprint = "1711.02180",
    archivePrefix = "arXiv",
    primaryClass = "hep-ph",
    reportNumber = "PITT-PACC-1712, IPPP-17-74, COEPP-MN-17-17",
    doi = "10.3389/fphy.2018.00040",
    journal = "Front. in Phys.",
    volume = "6",
    pages = "40",
    year = "2018"
}

@article{akhmedov2007interplay,
  author = {{Akhmedov}, Evgeny Kh. and {Frigerio}, Michele},
        title = "{Interplay of type I and type II seesaw contributions to neutrino mass}",
      journal = {Journal of High Energy Physics},
         year = 2007,
        month = jan,
       volume = {2007},
       number = {1},
          eid = {043},
        pages = {043},
          doi = {10.1088/1126-6708/2007/01/043},
archivePrefix = {arXiv},
       eprint = {hep-ph/0609046},
 primaryClass = {hep-ph},
       adsurl = {https://ui.adsabs.harvard.edu/abs/2007JHEP...01..043A}
}
\end{document}